\documentclass[aps,prb,reprint,groupedaddress,amsmath,amssymb,showpacs]{revtex4-1}

\usepackage[colorlinks=true,allcolors=blue,breaklinks]{hyperref}
\usepackage{graphicx}
\usepackage{dcolumn}
\usepackage{bm}
\usepackage{nicefrac}
\usepackage{stackengine}
\usepackage{amsmath}

\def\shrinkage{2.1mu}
\def\vecsign{\mathchar"017E}
\def\dvecsign{\smash{\stackon[-1.95pt]{\mkern-\shrinkage\vecsign}{\rotatebox{180}{$\mkern-\shrinkage\vecsign$}}}}
\def\dvec#1{\def\useanchorwidth{T}\stackon[-4.2pt]{#1}{\,\dvecsign}}
\stackMath

\begin{document}

\title{Two-dimensional quantum-spin-1/2 XXZ magnet in zero magnetic field: \\global thermodynamics from renormalization group theory}
\author{Ozan S. Sar{\i}yer}
\email[]{ossariyer@pirireis.edu.tr}
\affiliation{P\^{\i}r\^{\i} Reis University, Istanbul 34940 Turkey}
\date{\today}

\begin{abstract}
Phase diagram, critical properties and thermodynamic functions of the two-dimensional field-free quantum-spin-1/2 XXZ model has been calculated globally using a numerical renormalization group theory. The nearest-neighbor spin-spin correlations and entanglement properties, as well as internal energy and specific heat are calculated globally at all temperatures for the whole range of exchange interaction anisotropy, from XY limit to Ising limits, for both antiferromagnetic and ferromagnetic cases. We show that there exists long-range (quasi-long-range) order at low-temperatures, and the low-lying excitations are gapped (gapless) in the Ising-like easy-axis (XY-like easy-plane) regime. Besides, we identify quantum phase transitions at zero-temperature.
\end{abstract}

\pacs{05.10.Cc 
	05.30.Rt 
	05.50.+q 
	64.60.ae 
	64.60.Cn 
	64.60.De 
	64.70.Tg 
	75.10.Jm 
	75.10.Kt 
	75.30.Gw 
	75.30.Kz 
	75.40.Cx 
}

\maketitle

\section{Introduction}
Two-dimensional ($d=2$) quantum spin lattice models have attracted much attention, mainly due to the presence of magnetic monolayers in high-$T_c$ superconductors ~\cite{Anderson87}, while superfluid films have also been related to two-dimensional quantum magnetism \cite{Matsubara56, Matsuda57}. The isotropic Heisenberg (XXX) model has an $SU(2)$ symmetry, which remains unbroken at finite temperatures for $d\leqslant2$, as suggested by the Mermin-Wagner theorem. \cite{Mermin66} The model does not show a finite-temperature phase transition in $d=2$, \cite{Mermin66, Yamaji73} unless a symmetry-breaking external magnetic field or an interaction anisotropy is present \cite{Lines67, Plascak84, Yunoki02, Kar17, Ding90b}. As shall be discussed in Section~\ref{sec:PhaseDiag}, even a slight anisotropy can induce an ordered phase at finite-temperatures.

A particular case is the uniaxial magnetic anisotropy, which can be easy-axis or easy-plane in real materials, for which the axial and planar components of neighboring spins interact with different exchange interaction parameters as modeled by the anisotropic Heisenberg (XXZ) model. This type of magnetic anisotropy results from crystal field (due to lattice distortions) and spin-orbit coupling in magnetic materials, like in Ba$_3$CoSb$_2$O$_9$, for which the effective-spin-$\nicefrac{1}{2}$ Co$^{2+}$ ions form monolayers, with easy-plane-type anisotropic intralayer interactions (and negligibly weak interlayer interactions). \cite{Ding90b, Ding92b, Zhou12, Susuki13, Shirata12, Koutroulakis15, Marmorini16, Yamamoto17}

We use the spin-spin interactions language throughout the paper. But the quantum spin degrees of freedom appearing in the XXZ Hamiltonian [see eq.~(\ref{eq:XXZ1})], need not actually correspond to physical spins of atoms on a crystal lattice. One can interpret $s_i^z$ as an occupation operator, with eigenstates $\left|\uparrow\right\rangle$ and $\left|\downarrow\right\rangle$ corresponding to occupied and empty lattice site $i$. In this case, the $z$-interaction $s_i^z s_j^z$ models the attractive (repulsive) potential energy between nearest-neighbors for $\Delta>0$ ($\Delta<0$) [see eq.~(\ref{eq:XXZ1})]. Furthermore, the $xy$-interactions $\left(s_i^xs_j^x+s_i^ys_j^y\right)$ can be written using the spin ladder operators as $\frac{1}{2}\left(s_i^+s_j^-+s_i^-s_j^+\right)$, which is analogous to nearest-neighbor hopping kinetic energy term for particles. \cite{Yunoki02, Yang66a, Matsubara56, Matsuda57}

When written in this language of interacting hard-core bosons (or spinless fermions, or magnons) under the Matsubara-Matsuda transformation, the XXZ Hamiltonian models superfluids \cite{Matsubara56, Matsuda57}, supersolids \cite{Melko05, Melko07}, striped supersolids \cite{Melko06}, and valance-bond solids \cite{Isakov06}. The XXZ Hamiltonian in $d=2$ dimensions can be used to model not only two-dimensional magnetic crystals like K$_2$CuF$_4$ \cite{Sachs13}, V$X_2$ ($X$ = S, Se) \cite{Ma12}, $M$P$X_3$ ($M$ = V, Cr, Mn, Fe, Co, Ni, Cu, Zn, and $X$ = S, Se, Te) \cite{Chittari16, Lee16}, and magnetic monolayers in materials like CrI$_3$ \cite{McGuire15, Lado17} and in high-$T_\text{c}$ superconductors like La$_2$CuO$_4$ \cite{Ding90b, Makivic91, Anderson87, Manousakis91, Barnes91, Aplesnin98}, but also to model other systems exhibiting topological excitations, such as superfluid films, lipid layers \emph{etc}. \cite{Lee05}.

The quantum spin lattice models, like the XXZ model, can be experimentally simulated by artificially designed quantum systems of Rydberg atoms stored in magnetic microtraps \cite{Whitlock17}, by lattice constructions using low-temperature scanning tunneling microscopy \cite{Toskovic16}, by trapped ion-laser systems \cite{Richerme14, Jurcevic14, Grass15, Nath15}, and by ultracold bosonic atoms in optical lattices \cite{Fukuhara13a, Fukuhara13b}. Such systems are considered to be important in realizations of quantum computers and spintronic devices. \cite{Loss98, Burkard99, Glaser03, Amico08}

Despite the ever increasing interest, the exact solution for the XXZ model at finite temperatures in $d>1$ dimensions is missing. Investigations on the model usually cover specific regimes, such as high-temperature or weak-anisotropy. The lack of an exact solution is basically due to the non-commutativity of spin operators between nearest-neighboring sites. Suzuki and Takano proposed an approximate renormalization group (RG) method that essentially neglects this non-commutativity, and they obtained the phase diagram and critical properties for the model in $d=2$ and $3$ dimensions. \cite{Suzuki79, Takano81}

Previously, we used the Suzuki-Takano approach to calculate thermodynamic functions of the XXZ model in $d=1$ dimensions, and showed that the approximate RG procedure works well even in the low-temperature regime. \cite{Sariyer08} In this article, we study the model in $d=2$ dimensions. We reproduce the results of Suzuki and Takano (phase diagram and critical properties), and extend their work by calculating the thermodynamic functions globally at all temperatures and anisotropies. We obtain numerical results for the nearest-neighbor spin-spin correlations and entanglement properties, besides internal energy and specific heat. We show the existence of long-range (quasi-long-range) order at low-temperatures, and that the low-lying excitations are gapped (gapless) in the Ising-like easy-axis (XY-like easy-plane) regime.

Although the Suzuki-Takano RG method is essentially a high-temperature approximation, we still obtain good results at low-temperatures, which compare at least qualitatively well with the results obtained by other methods. We can even identify the quantum phase transitions at zero-temperature, where the method is expected to work at its worst. The advantage here is the capability of capturing global thermodynamics by a single method that does not require excess computational power. Hence we expect our results to shed light on the thermodynamic and entanglement properties of the real systems that possess uniaxial magnetic anisotropy. In the following, we will first introduce the model (Section~\ref{sec:model}) and the methods (Section~\ref{sec:methods}), and then discuss our results (Section~\ref{sec:results}) before the conclusion (Section~\ref{sec:conclusion}).

\section{XXZ model}\label{sec:model}
The spin-$\nicefrac{1}{2}$ anisotropic quantum Heisenberg model (XXZ model) is defined by the Hamiltonian
\begin{equation}
\label{eq:XXZ1}
\mathcal{H}=-\tilde{J}\frac{3}{2+\left|\Delta\right|}\sum_{\langle ij\rangle}\left[s_i^xs_j^x+s_i^ys_j^y+\Delta s_i^zs_j^z\right],
\end{equation}
where the sum is over ``$ij$-bonds'', \emph{i.e.}, over nearest-neighboring lattice sites $i$ and $j$. The operators $s_i^u$ with $u\in\left\{x,y,z\right\}$ are the quantum mechanical spin-$\nicefrac{1}{2}$ ($S=1/2$) operators acting at site $i$ (with $s_i^z$ eigenvalues $\pm\nicefrac{1}{2}$ for the factor $\hbar^2$ absorbed in $\tilde{J}$) that obey the commutation relation $\left[s_i^u,s_j^v\right]=i\delta_{ij}\epsilon_{uvw}s_i^w$. The model becomes classical in the limit $S\to\infty$, while quantum mechanical effects are maximal for the smallest possible spin value, $S=1/2$.

The anisotropy parameter $\Delta$ in XXZ Hamiltonian~(\ref{eq:XXZ1}) (ratio of $s^z$-interactions to $s^x$- or $s^y$-interactions) makes the Hamiltonian interpolate continuously between classical Ising, quantum XXX, and quantum XY models. For positive $\tilde{J}$, the quantum XXZ model reduces to classical ferromagnetic (FM) and antiferromagnetic (AFM) Ising models in the limits $\Delta\to\infty$ and $\Delta\to-\infty$ respectively. At $\Delta=1$ ($\Delta=-1$), the anisotropic XXZ model Hamiltonian~(\ref{eq:XXZ1}) reduces to the isotropic FM (AFM) XXX model Hamiltonian, while at $\Delta=0$, it turns into the quantum XY (or XX0) model Hamiltonian. We call the regimes of $\left|\Delta\right|>1$ and $\left|\Delta\right|<1$, the ``Ising-like'' regime and the ``XY-like'' regime respectively. These regimes model materials that respectively possess easy-axis and easy-plane magnetic anisotropies.

The parameter $\tilde{J}$ in Hamiltonian~(\ref{eq:XXZ1}) is the material-dependent exchange interaction energy between nearest-neighbor spins. We define a temperature-dependent dimensionless interaction parameter
\begin{equation}
\label{eq:J}
J\equiv\beta\tilde{J}\equiv\frac{\tilde{J}}{kT}\equiv\frac{\tilde{T}}{T},
\end{equation}
where, $kT\equiv\beta^{-1}$ is the thermal energy (Boltzmann constant $k$ times absolute temperature $T$), and $\tilde{T}\equiv\tilde{J}/k$ is the temperature scale associated with the material.

For a proper application of RG theory, we rewrite the Hamiltonian~(\ref{eq:XXZ1}) in a dimensionless form as
\begin{equation}
\label{eq:XXZ}
-\beta\mathcal{H}=\sum_{\langle ij\rangle}\left[J_x\left(s_i^xs_j^x+s_i^y s_j^y\right)+J_zs_i^z s_j^z+G\right].
\end{equation}
Here, we defined the dimensionless exchange interaction parameters $J_x\equiv\frac{3J}{2+\left|\Delta\right|}=\frac{3}{2+\left|\Delta\right|}\frac{\tilde{T}}{T}$ and $J_z\equiv\frac{3J\Delta}{2+\left|\Delta\right|}=\frac{3\Delta}{2+\left|\Delta\right|}\frac{\tilde{T}}{T}=\Delta J_x$. From the definition~(\ref{eq:J}), we rewrite the temperature as
\begin{equation}
\label{eq:T}
\frac{T}{\tilde{T}}=\frac{1}{J}=\frac{3}{2J_x+\left|J_z\right|}.
\end{equation}
Scaling all temperatures by a factor $\tilde{T}$ ensures a universality between different materials of same anisotropy parameter $\Delta$. The factor $\frac{3}{2+\left|\Delta\right|}$ in Hamiltonian~(\ref{eq:XXZ1}), ensures the same definition~(\ref{eq:T}) of a scaled temperature for the whole range of $\Delta$, even in the Ising model limits.

In writing the Hamiltonian~(\ref{eq:XXZ}), we introduced a dimensionless parameter $G$. This term shifts the dimensionless energy of the system by a \emph{constant} amount, $G$ per bond, and thus, does not affect the thermodynamics of the system. We can take $G=0$ for an original system. However, when we renormalize the system, this parameter maps to a positive $G^\prime>0$, which accounts for the entropy of fluctuations of the renormalized degrees of freedom. Here, and throughout the article, we use primes to denote the interaction parameters and thermodynamic densities of the renormalized system.

Due to the axial $U(1)$ symmetry of the Hamiltonian under $x\leftrightarrow y$, all thermodynamic quantities associated with the $y$-direction are the same as those associated with the $x$-direction. In example, the nearest-neighbor spin-spin correlations obey $\left\langle s_i^ys_j^y\right\rangle=\left\langle s_i^xs_j^x\right\rangle$ at all temperatures and all anisotropies. Hence, we omit the direction-$y$ for the rest of the paper.

\section{Methods}\label{sec:methods}
	\subsection{Renormalization group transformation for \textit{d}\,=\,1} \label{sec:RG1d}
We use the approximate RG theory developed by Suzuki and Takano for quantum lattice systems in $d=1$ dimensions \cite{Suzuki79, Takano81}. An RG transformation with rescaling factor $b=2$, maps the original system of $N$ original spins with original interaction parameters $\vec{K}=\left(J_x,J_z,G\right)$ onto a renormalized system of $N^\prime=N/b=N/2$ spins with renormalized interaction parameters $\vec{K}^\prime=\left(J_x^\prime,J_z^\prime,G^\prime\right)$. Under the RG transformation, the form of the Hamiltonian~(\ref{eq:XXZ}) stays the same, and in the thermodynamic limit (in particular for $N\to\infty$ with periodic boundary conditions), the partition function (and hence the free energy of the system) stays invariant. This RG transformation of real-space rescaling factor $b=2$ is carried out by integration over every other spin (say spins at odd-$j$ sites).

In Suzuki-Takano approach for $d=1$, this $b=2$ renormalization procedure is approximately formulated as \cite{Suzuki79, Takano81}
\begin{equation}
\label{eq:ST}
e^{-\beta^\prime\mathcal{H}_{ik}^\prime}=\text{Tr}_j\,e^{-\beta\mathcal{H}_{ij}-\beta\mathcal{H}_{jk}},
\end{equation}
where $i$, $j$, $k$ are three successive sites in one-dimensional lattice, and $\beta\mathcal{H}_{ij}$ is the dimensionless Hamiltonian operating on the $ij$-bond, such that the system Hamiltonian~(\ref{eq:XXZ}) reads $\beta\mathcal{H}=\sum_{\langle ij\rangle}\left[\beta\mathcal{H}_{ij}\right]$.\cite{Sariyer08} In equation~(\ref{eq:ST}), the operator $-\beta^\prime\mathcal{H}_{ik}^\prime$ acts on two-site states $\left|s_i\right>\otimes\left|s_k\right>$ of the renormalized system, while the operator $-\beta\mathcal{H}_{ij}-\beta\mathcal{H}_{jk}$ acts on three-site states $\left|s_i\right>\otimes\left|s_j\right>\otimes\left|s_k\right>$ of the original system.

In Suzuki-Takano RG approximation~(\ref{eq:ST}), we assume vanishing commutators $\left[-\beta\mathcal{H}_{ij},-\beta\mathcal{H}_{mn}\right]=0$ for $\left|m-i\right|\geqslant b$. Although the non-commutativity of operators are neglected only beyond $b+1$ successive sites, for the XXZ model, only the nearest-neighbor operators do not commute with each other, and we always neglect the non-commutativity of the operators at the two ends of the RG clusters of size $b$. Hence, choosing a larger $b$ would not improve the Suzuki-Takano approximation for the XXZ model.

This approximation works best at high temperatures (small $J$), since the neglected commutators are $\mathcal{O}\left(e^{\left[-\beta\mathcal{H}_{ij},-\beta\mathcal{H}_{mn}\right]}  \right)=\mathcal{O}\left(e^{J^2}\right)$ at the first order of Baker-Campbell-Hausdorff formula. Moreover, since the same approximation is applied in factorizing $e^{-\beta\mathcal{H}}$ and $e^{-\beta^\prime\mathcal{H}^\prime}$ in opposite directions, the first order corrections, $\mathcal{O}\left(e^{J^2}\right)$ and $\mathcal{O}\left(e^{-J^2}\right)$, are expected to roughly cancel each other out \cite{Suzuki79, Tatsumi81}, and the leading order corrections become $\mathcal{O}\left(e^{J^4}\right)$ \cite{Takano81, Tatsumi81}. Hence, even at zero-temperature limit ($J\to\infty$), results for thermodynamic functions obtained for $b=2$ and $d=1$, compare well with exact results. \cite{Sariyer08}

On the anisotropy axis, the approximation becomes exact at the Ising limits, $\left|\Delta\right|\to\infty$, where the operators become classical and commute with each other. Hence, we expect the worst results for the XY model ($\Delta=0$) at zero-temperature. The nature of the approximation in this limit for the \emph{critical behavior} in the XY-like regime has been discussed in detail. \cite{Takano81, Barma79} In this article we apply the Suzuki-Takano approach to calculate \emph{thermodynamic functions} of the XXZ model in $d=2$, and again we expect the worst results in the limit $\Delta=0$ and $T=0$. In example, there is a $20\%$ discrepancy between the ground-state energy we calculate for the XY model and the square-lattice results by various methods (see Section~\ref{sec:ZeroT}). Even so, we still obtain qualitatively good results for the global ranges of temperature and anisotropy.

In $d=1$ dimensions, previously we have obtained the recursion relations between the renormalized (\emph{i.e.}, $J_x^\prime$, $J_z^\prime$, $G^\prime$) and the original (\emph{i.e.}, $J_x$, $J_z$, $G$) interaction parameters via equation~(\ref{eq:ST}) as \cite{Takano81, Sariyer08}
\begin{eqnarray}
\nonumber J_x^\prime&=&\ln\left[e^{-J_z/4}\left(\cosh\bar{J}+\frac{J_z}{4\bar{J}}\sinh\bar{J}\right)\right],\\
\label{eq:recursion} J_z^\prime&=&\ln\left[\frac{e^{-J_z/4}\left(e^{3J_z/4}+\cosh\bar{J}-\frac{J_z}{4\bar{J}}\sinh\bar{J}\right)^2}{4\left(\cosh\bar{J}+\frac{J_z}{4\bar{J}}\sinh\bar{J}\right)}\right],\\
\nonumber G^\prime&=&2G+\frac{2J_x^\prime+J_z^\prime}{4}+\ln2,
\end{eqnarray}
where we defined $\bar{J}\equiv\frac{1}{4}\sqrt{8J_x^2+J_z^2}$ for simplicity. One expected result is the dependence of the renormalized parameters on the additive parameter $G$, \emph{i.e.}, $\partial_GJ_x^\prime=\partial_GJ_z^\prime=0$ and $\partial_GG^\prime=b^d=2$. Another expected result is that the recursion relations~(\ref{eq:recursion}) are invariant under a sign change of $J_x$. In fact, this invariance is a special case of a more general symmetry of the XYZ model. In this more general model, the Hamiltonian operator is symmetric under sign changes of two of the three interaction coefficients $J_x$, $J_y$, and $J_z$. \cite{Yang66c} Note that $J_y=J_x$ in the XXZ model Hamiltonian~(\ref{eq:XXZ}). Due to this sign symmetry $J_x\leftrightarrow-J_x$, in the following, we consider only the $J_x\geqslant0$ subspace of the interaction parameters.

	\subsection{Renormalization group transformation for \textit{d}\,\textgreater\,1}
The recursion relations~(\ref{eq:recursion}) are in the form $\vec{K}^\prime=R(\vec{K})$. These recursion relations for a one-dimensional system can be generalized to those for an arbitrary $d$-dimensional system, by using the Migdal-Kadanoff RG procedure as \cite{Migdal75, Kadanoff76}
\begin{equation}
\label{eq:MK}
\vec{K}^\prime=R\left(b^{d-1}\vec{K}\right).
\end{equation}
We should note that while the original Suzuki-Takano approach \cite{Suzuki79, Takano81} applies one-dimensional decimation followed by bond-moving, \emph{i.e.}, $\vec{K}^\prime=b^{d-1}R(\vec{K})$, here we apply the opposite: bond-moving followed by decimation, \emph{cf.} equation~(\ref{eq:MK}).

The Migdal-Kadanoff approach applies to classical models, as well as to quantum models \cite{Tatsumi81}, on the same bond-moving basis due to Hermiticity of the Hamiltonians \cite{Takano81}. The Migdal-Kadanoff approximation becomes exact at infinite-temperature ($J=0$), since bond-moving has no effect in this limit. Recalling that the Suzuki-Takano approach also works best at high-temperatures for $d=1$, for quantum models in $d>1$ at finite-temperatures, we cannot separate the errors due to Migdal-Kadanoff approximation (bond-moving) from those due to Suzuki-Takano approximation (neglecting non-commutativity) \cite{Tatsumi81}.

\begin{figure}
\includegraphics[scale=1.0]{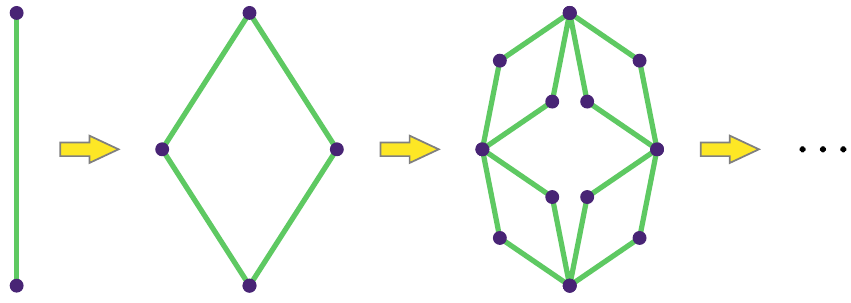}
\caption{\label{fig:Hierarchical} Construction of the hierarchical lattice for $b=2$ and $d=2$. At each step, every bond is replaced by two parallel paths ($d=2$) each of which contains two bonds in series ($b=2$). Repeating the process \emph{ad infinitum}, yields the two-dimensional hierarchical lattice, on which the Migdal-Kadanoff RG equation~(\ref{eq:MK}) works exact for classical models.}
\end{figure}

Migdal-Kadanoff RG procedure works exact for classical models in $d$-dimensional hierarchical lattices, while it can be considered as a good approximation for any other $d$-dimensional lattice (\emph{e.g.} hypercubic lattices), especially for classical Hamiltonians. \cite{Berker79, Kaufman81, Griffiths82, Kaufman84, Erbas05} The hierarchical lattice we use here for $b=2$ and $d=2$ is presented in Fig.~\ref{fig:Hierarchical}. We obtain the recursion relations using the one-dimensional relations~(\ref{eq:recursion}) in equation~(\ref{eq:MK}) with $b=2$ and $d=2$. This approach has been used to study superfluid systems \cite{Sano87, Doi89}, as well as electronic systems such as Hubbard \cite{Cannas91, Cannas92, Migliorini00, Hinczewski05}, $t$-$J\,$ \cite{Falicov95, Hinczewski06, Hinczewski08, Kaplan09} and Falicov-Kimball \cite{Sariyer11} models.

	\subsection{Calculation of the phase diagram}
From the RG flow diagram (see Fig.~\ref{fig:FlowDiag}), we can calculate the phase diagram of the system (see Fig.~\ref{fig:PhaseDiag}). Under successive RG transformations, a point $\vec{K}$ in the interaction parameters space flows to a sink. In the anisotropic XXZ model, the RG flows happen in the $J_xJ_zG$-space. Under successive RG transformations, the parameter $G$ always grows to infinity, since integrating more and more spins into a single renormalized spin adds more and more entropy associated with the fluctuations of the integrated spin degrees of freedom. We take $G=0$ for an original system, and in the $J_xJ_z$-plane, different phases are characterized by flows to different sinks (see Table~\ref{tab:sinks}). Each transition between different phases is controlled by a corresponding fixed point (see Table~\ref{tab:criticalpoints}). This critical fixed point determines the universality class of the transition.

	\subsection{Calculation of thermodynamic functions}
		\subsubsection{Calculation of spin-spin correlations}
For each type of interaction $K_\alpha$ appearing in a model Hamiltonian, we can define a density (\emph{i.e.}, the expectation value of the operator associated with $K_\alpha$) as
\begin{equation}
M_\alpha=\frac{1}{N_\alpha}\frac{\partial\ln Z}{\partial K_\alpha}.
\end{equation}
Here, $Z$ is the partition function ($\ln Z=-\beta F$ is the dimensionless negative Helmholtz free energy), and $N_\alpha$ is the number of $\alpha$-type interactions. The interactions in XXZ Hamiltonian~(\ref{eq:XXZ}) all act through bonds, and hence, all $N_\alpha$ are the same and equal to the number of bonds in the system. These interactions are $\vec{K}=\left(J_x,J_z,G\right)$ and the corresponding densities are $\vec{M}=\left(2\langle s_i^xs_j^x\rangle,\langle s_i^zs_j^z\rangle,1\right)$. Here, $\langle s_i^xs_j^x\rangle=\langle s_i^ys_j^y\rangle$ and $\langle s_i^zs_j^z\rangle$ are the averages of the nearest-neighbor spin-spin correlations. The unit operator associated with the parameter $G$ has the constant eigenvalue $1$, which has the constant average $1$.

As the interaction parameters of the renormalized and original systems ($\vec{K}^\prime$ and $\vec{K}$) are connected by recursion relations, the densities in renormalized and original systems ($\vec{M}^\prime$ and $\vec{M}$) are connected by the recursion matrix $\dvec{T}$ as \cite{Sariyer08, McKay84}
\begin{equation}
\label{eq:densities}
b^d\vec{M}=\vec{M}^\prime\cdot\dvec{T}.
\end{equation}
Elements of the recursion matrix are $T_{\gamma\alpha}=\frac{N_\gamma}{N_\alpha}\frac{\partial K_\gamma^\prime}{\partial K_\alpha}$. For the XXZ model Hamiltonian~(\ref{eq:XXZ}), since all $N_\alpha$ are the same, we have $T_{\gamma\alpha}=\frac{\partial K_\gamma^\prime}{\partial K_\alpha}$.

At a fixed point such as a sink, RG transformation keeps the interaction parameters and hence the densities invariant: $\vec{M}=\vec{M}^\prime=\vec{M}^\ast$. Therefore, at a sink, relation~(\ref{eq:densities}) takes the form of a left-eigenvalue equation:
\begin{equation}
\label{eq:lefteigen}
b^d \vec{M}^\ast=\vec{M}^\ast\cdot\dvec{T}.
\end{equation}
To calculate densities vector $\vec{M}$ at an ordinary point (rather than a fixed point), we iterate equation~(\ref{eq:densities}) $n$-times, until we get sufficiently close to a sink point, where the densities are approximately $\vec{M}^\ast$. Hence we obtain the densities $\vec{M}$ of the original system iteratively as \cite{Sariyer08, McKay84}
\begin{equation}
\label{eq:original}
\vec{M}=b^{-nd}\vec{M}^\ast\cdot\dvec{T}^{(n)}\cdot\dvec{T}^{(n-1)}\cdots\dvec{T}^{\prime\prime}\cdot\dvec{T}^\prime\cdot\dvec{T}.
\end{equation}
We can calculate $\vec{M}^\ast$ from the eigenvalue equation~(\ref{eq:lefteigen}), which, substituting into equation~(\ref{eq:original}), yields the densities $\vec{M}$ at the original point.

\begin{figure*}[t!]
	\includegraphics[scale=1.0]{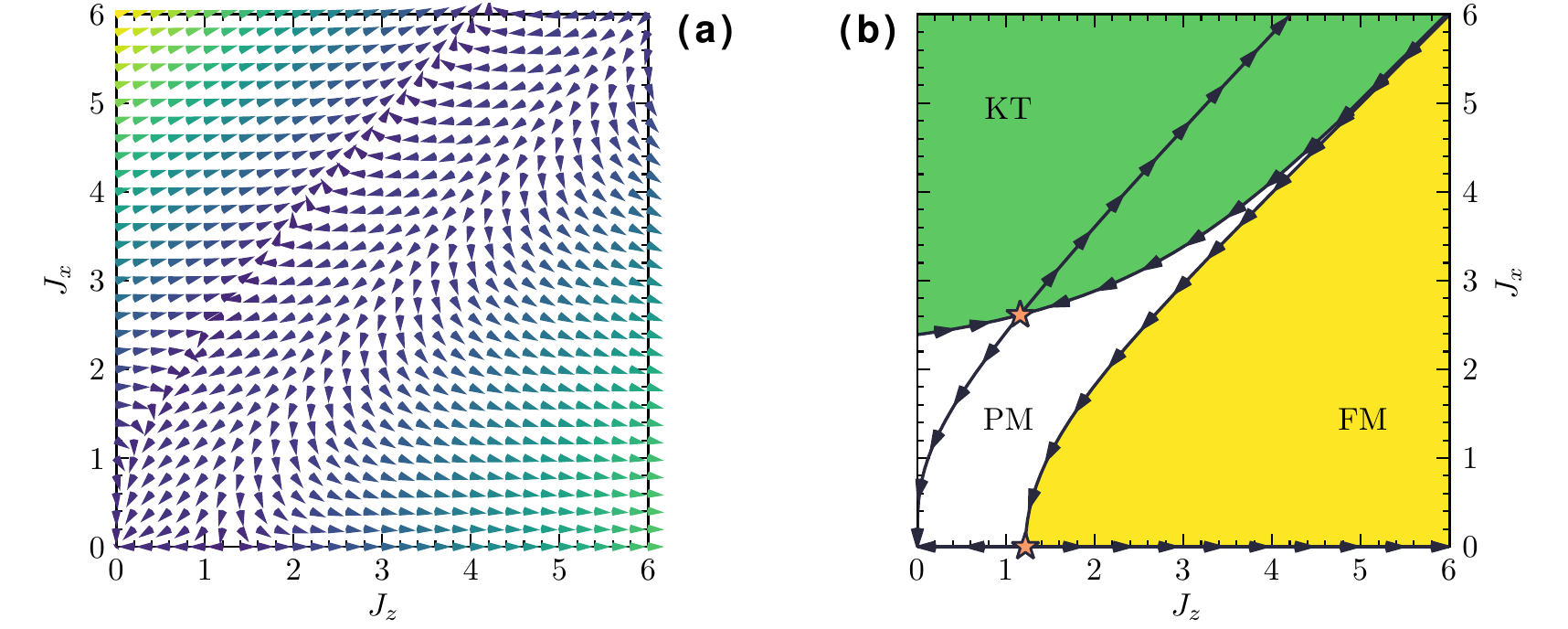}
	\caption{\label{fig:FlowDiag} RG flows in $J_zJ_x$-space for $J_z>0$ and $J_x>0$. Arrows denote directions of RG flows in both panels. Panel (a): Fine detailed flow field. Colors from purple to yellow represent the speed of RG flows, $\sqrt{(J_z^\prime-J_z)^2+(J_x^\prime-J_x)^2}$, from slow to fast. Note the expected slowing down of RG flows close to critical points. Panel (b): Phases and critical lines. All points in white region flow to infinite-temperature paramagnetic (PM) phase sink located at $J_z=0$, $J_x=0$; in yellow region to zero-temperature ferromagnetic (FM) phase sink at $J_z\to\infty$, $J_x=0$; and in green region to zero-temperature Kosterlitz-Thouless (KT) phase sink at $J_z\to\infty$, $J_x\to\infty$ (see Table~\ref{tab:sinks}). The two stars denote Ising and XY fixed points controlling the thermodynamics of (A)FM-PM and KT-PM phase transitions (see Table~\ref{tab:criticalpoints}).}
\end{figure*}

		\subsubsection{Calculation of internal energy}
From the densities vector, we can extract the nearest-neighbor spin-spin correlations $\langle s_i^us_j^u\rangle$, and calculate the dimensionless internal energy density $U/\tilde{J}$ of the system (internal energy per bond in units of $\tilde{J}$) as
\begin{equation}
\label{eq:Uint}
\frac{U}{\tilde{J}}=\frac{\langle\mathcal{H}_{ij}\rangle}{\tilde{J}}=\frac{-3}{2+\left|\Delta\right|}\left(2\langle s_i^xs_j^x\rangle+\Delta\langle s_i^zs_j^z\rangle\right) .
\end{equation}
Internal energy density calculated in the literature with different methods, can be trivially related to our results obtained by equation~(\ref{eq:Uint}). In example, one commonly adopted form of the dimensionless XXZ Hamiltonian in the literature is $\bar{\mathcal{H}}=-\sum_{\langle ij\rangle}\left[\sigma_i^x\sigma_j^x+\sigma_i^y\sigma_j^y+\Delta\sigma_i^z\sigma_j^z\right]$, where $\sigma_i^u=2s_i^u$ are Pauli spin matrices with eigenvalues of $\sigma_i^z$ being $\pm1$. One can calculate the dimensionless internal energy \emph{per site} from this Hamiltonian as $\bar{U}=\frac{z}{2}\langle\bar{\mathcal{H}}_{ij}\rangle=-\frac{z}{2}\left(2\langle\sigma_i^x\sigma_j^x\rangle+\Delta\langle\sigma_i^z\sigma_j^z\rangle\right)=-2z\left(2\langle s_i^xs_j^x\rangle+\Delta\langle s_i^zs_j^z\rangle\right)$, where $z$ is the number of nearest-neighbors of a site, \emph{e.g.}, $z=2d$ for $d$-dimensional hypercubic lattice. Comparing with the internal energy \emph{per bond} of equation~(\ref{eq:Uint}), we simply relate the two internal energy densities with $U/\tilde{J}=\frac{3}{2z\left(2+\left|\Delta\right|\right)}\bar{U}$.

		\subsubsection{Calculation of specific heat}
Similar trivial connections can be deduced for temperature, anisotropy parameter, and the specific heat. From the internal energy density~(\ref{eq:Uint}), we can calculate the dimensionless specific heat $C/k$ (heat capacity per bond in units of $k$) as
\begin{equation}
\label{eq:Csph}
\frac{C}{k}=\frac{1}{k}\frac{\partial U}{\partial T}=\frac{\partial(U/\tilde{J})}{\partial(T/\tilde{T})}.
\end{equation}
We calculate densities at different temperatures separated by a small difference $d(T/\tilde{T})=10^{-3}$, and use a two-point numerical derivation procedure to calculate the specific heat as a function of temperature. We observe that employing a higher-accuracy numerical scheme (\emph{e.g.}, five-point numerical derivation) does not affect our numerical results, for our choice of $d(T/\tilde{T})=10^{-3}$ being sufficiently small, except at low temperatures (see Section~\ref{sec:LowT}), where we employed $d(T/\tilde{T})=10^{-6}$.

\section{Results and discussion}\label{sec:results}
	\subsection{Critical properties and the phase diagram}
		\subsubsection{RG flows, sinks and critical points}
In Fig.~\ref{fig:FlowDiag}, we plot RG flows obtained using relations~(\ref{eq:recursion}) and~(\ref{eq:MK}) with $b=2$ and $d=2$, which compares well with the flow diagram obtained by another approximate RG procedure by Dekeyser \emph{et al.} \cite{Dekeyser77}. Here, we plot flows only in the first quadrant of the $J_zJ_x$-space ($J_z>0$ and $J_x>0$) for graphical simplicity. Recall that we do not consider the third and fourth quadrants ($J_x<0$) of the $J_zJ_x$-space at all, due to the aforementioned $J_x\leftrightarrow-J_x$ symmetry (see Section~\ref{sec:RG1d}). Moreover, any point in the second quadrant of the $J_zJ_x$-space ($J_z<0$ and $J_x>0$), maps to a point in the first quadrant by a single RG transformation, and continues to flow in the first quadrant under further transformations. This is indeed an expected result since a three-site AFM state $\left|\uparrow\downarrow\uparrow\right\rangle$ maps onto a two-site FM state $\left|\uparrow\,\,\,\uparrow\right\rangle$ under a single RG transformation.

One interesting result is that while the classical Ising model line ($J_x=0$) and the isotropic quantum XXX model line ($J_x=J_z$) are closed under RG transformations, the quantum XY model line ($J_z=0$) is not: a point on the $J_z=0$ line maps to a finite $J_z^\prime>0$ under a single RG transformation step (see Figure \ref{fig:FlowDiag}(a)). This is because the XY fixed point is not on the $J_z=0$ line, but at a finite $J_z>0$ (see Figure \ref{fig:FlowDiag}(b)). This might be an error due to Suzuki-Takano approach as discussed previously. \cite{Tatsumi81, Barma79} Recall that the Suzuki-Takano approach is exact in the classical Ising limits ($\left|\Delta\right|\to\infty$), and we expect this approximation to worsen as we approach to XY model ($\left|\Delta\right|\to0$).

\begin{table}[t!]
	\caption{\label{tab:sinks}Characteristics at phase sinks: interaction parameters $J_u$, runaway coefficients $J_u^\prime/J_u$, and densities $\langle s_i^us_j^u\rangle$.}
	\begin{ruledtabular}
		\begin{tabular}{c c c c c c c}
			Sinks & $J_x$ & $J_z$ & $J_x^\prime/J_x$ & $J_z^\prime/J_z$ & $\langle s_i^xs_j^x\rangle$ & $\langle s_i^zs_j^z\rangle$ \\
			\hline
			PM & $0$ & $0$ & $0$ & $0$ & $0$ & $0$\\
			(A)FM & $0$ & $\infty$ & $0$ & $2$ & $0$ & $1/4$\\
			KT & $\infty$ & $\infty$ & $1$ & $1$ & $0$ & $1/4$
		\end{tabular}
	\end{ruledtabular}
\end{table}

\begin{table}[b!]
	\caption{\label{tab:criticalpoints}Critical interaction parameters $J_u^\ast$ and relevant critical exponents $y_T$ at Ising and XY critical fixed points (see star-markers in Fig.~\ref{fig:FlowDiag}(b)).}
	\begin{ruledtabular}
		\begin{tabular}{c c c c c}
			Transition & Universality class & $J_x^\ast$ & $J_z^\ast$ & $y_T$ \\
			\hline
			(A)FM-PM & Classical Ising & $0$ & $1.218756$ & $0.747236$\\
			KT-PM & Quantum XY & $2.612359$ & $1.156035$ & $0.164979$
		\end{tabular}
	\end{ruledtabular}
\end{table}

In Table~\ref{tab:sinks}, characteristics of the sink points are presented. Each sink point corresponds a different phase for the system. As mentioned above, the parameter $G$ always flows to infinity under successive RG transformations with a usual runaway coefficient $G^\prime/G=b^d=4$; and the unit operator associated with it has the constant expectation value $1$ at any point (not only at fixed points); hence they are not shown in Table~\ref{tab:sinks}.

Phase transitions are also characterized by fixed points. All the KT-PM ((A)FM-PM) transitions for $\left|\Delta\right|<1$ ($\left|\Delta\right|>1$) are in the quantum XY (classical Ising) universality class and are controlled by the XY (Ising) fixed point. We present the characteristics of phase transition lines in Table~\ref{tab:criticalpoints}. Here, $y_T=\log_b\lambda_T$ is the scaling exponent in the relevant direction, and $\lambda_T$ is the relevant eigenvalue of the recursion matrix $\dvec{T}$ at the critical fixed point [see equation~(\ref{eq:lefteigen})].

The critical interaction parameters $J_u^\ast$ (and also the relevant critical exponents $y_T$) in Table~\ref{tab:criticalpoints} are the same as those obtained by the original Suzuki-Takano approach with a different Migdal-Kadanoff scheme \cite{Suzuki79, Takano81}, and agree qualitatively with those obtained by the RG procedure developed by Dekeyser \emph{et al.}: finite $J_z^\ast=1.398$ with vanishing $J_x^\ast=0$ for the Ising fixed point, and finite $J_x^\ast=1.478$ with non-zero $J_z^\ast=0.064$ for the XY fixed point \cite{Dekeyser77}. However, we should note that this approach of Dekeyser \emph{et al.} also leads to a finite-temperature critical fixed point for the XXX model at $J_x^\ast=J_z^\ast=2.854$, which violates the rigorous Mermin-Wagner theorem  \cite{Mermin66}, \emph{i.e.}, the absence of a finite-temperature phase transition for XXX model in $d=2$.

Our calculated value of $y_T=0.747$ for (A)FM-PM transition is within $25\%$ error range of Onsager’s exact result $y_T=1$ for square lattice \cite{Onsager44}. We note that although the square lattice approximation by the hierarchical lattice in Fig. \ref{fig:Hierarchical} (or equivalently the Migdal-Kadanoff approximation) results in critical temperatures with small errors, it might lead to large errors in critical exponents. \cite{Tatsumi81} Note that other critical exponents --such as $\alpha=2-d/y_T$ and $\nu=1/y_T$ for the power-law dependences of specific heat $C\sim(T-T_c)^{-\alpha}$ and correlation length $\xi\sim(T-T_c)^{-\nu}$-- can be calculated from $y_T$ for the Ising transition. We also note that the calculated exponent $y_T$ for the KT-PM transition is smaller than that for the (A)FM-PM transition as expected \cite{Suzuki72}.

Although $y_T=0.165$ for the KT-PM transition seems quite small, we should note that the KT-PM transition is a Kosterlitz-Thouless transition \cite{Kosterlitz73, Kosterlitz74}, rather than a usual second-order transition. In particular, there is no symmetry breaking, and the thermodynamic functions have KT-type exponential singularities \cite{Kosterlitz74, Ding92b} (such as $C\sim e^{A(T-T_c)^{-\bar{\alpha}}}$) rather than power-law singularities. It was also argued that the hyperscaling relation may be violated for the XY universality class. \cite{Rogiers79} This XY universality applies to many diverse systems that undergo a KT phase transition in $d=2$, such as superfluids \cite{Bishop78}, superconductors \cite{Hebard80, Epstein81, Resnick81}, Josephson junction arrays \cite{Martinoli00}, Bose-Einstein condensates \cite{Hadzibabic06}, and two-dimensional solids \cite{Strandburg88}. In addition, a small value of $y_T$ for XY-like regime, translates into a weak singularity in specific heat (see Fig.~\ref{fig:UiCsAll}(b)), which is in accord with the non-singular specific heat results obtained by Monte Carlo simulations \cite{Ding92b, Rogiers79, Suzuki77, Loh85a, Loh85b, Ding90a, Ding92a, Capriotti97, Cuccoli03}. We also note that for the XY model, a large $y_T=1.300$ (for $b=\sqrt{2}$) and a small $y_T=0.650$ (for $b=2$) for $d=2$; and a marginal $y_T=0$ for a slightly larger $2.05<d<2.23$ were obtained by RG methods similar to ours. \cite{Tatsumi81} We should also recall that the Migdal-Kadanoff approximation is not so good in estimating critical exponents, and that the Suzuki-Takano approximation works at its worst in the XY-like regime ($|\Delta|<1$), where the KT-PM fixed point appears (at $\Delta^\ast=J_z^\ast/J_x^\ast=0.442525$, see Table~\ref{tab:criticalpoints}).
\begin{figure*}[t!]
	\begin{tabular}{c}
		\includegraphics[scale=1.0]{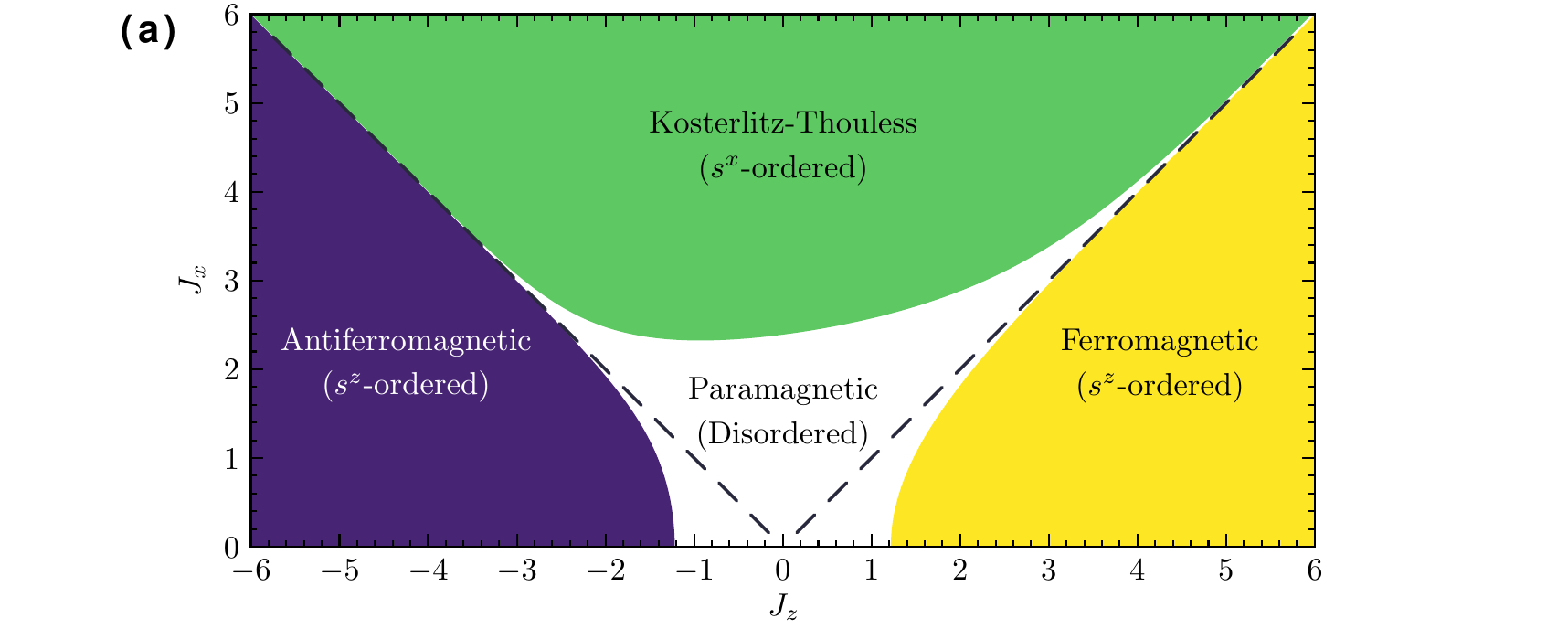}\\
		\includegraphics[scale=1.0]{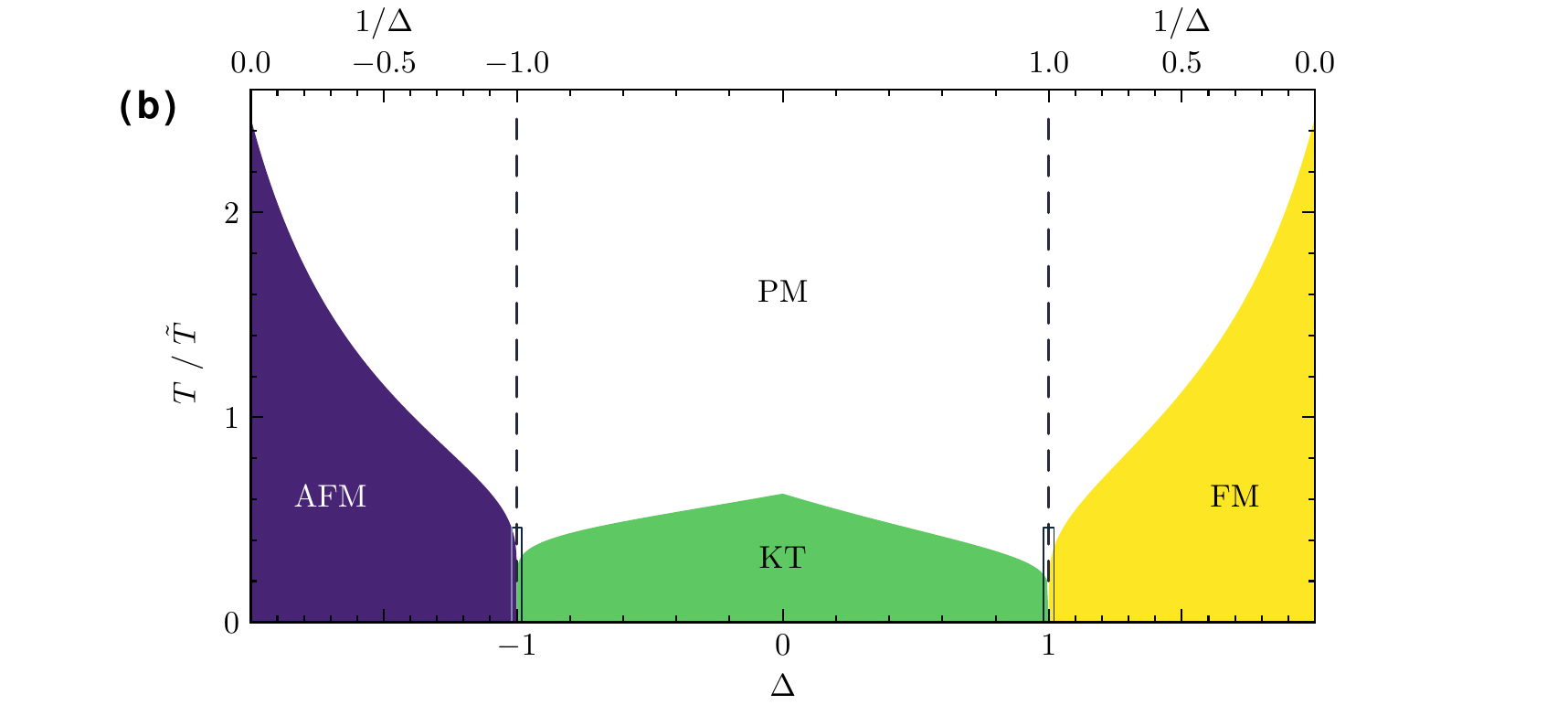}\\
	\end{tabular}
	\caption{\label{fig:PhaseDiag} Phase diagram of the two-dimensional quantum XXZ model in $J_zJ_x$-space (a) and in anisotropy-temperature space (b). In panel (b), we respectively employ $\Delta$ and $1/\Delta$ as the horizontal axis in XY-like regime ($|\Delta|<1$) and in Ising-like regimes ($|\Delta|>1$). In both panels, dashed lines are not phase boundaries, but just mark the AFM and FM isotropic XXX models ($J_x=|J_z|$, $|\Delta|=1$). The ordered phases meet only at the zero-temperature isotropic points ($J_x=|J_z|\to\infty$, $|\Delta|=1$, $T=0$), hence, there is no ordered phase for the isotropic XXX models for $T>0$. For the discussion on zero-point quantum phase transitions at $|\Delta|=1$ and $T=0$, see Section~\ref{sec:ZeroT}. For a zoomed view of the boxed regions of panel (b) (around $|\Delta|=1$), see Fig.~\ref{fig:Tc1}.}
\end{figure*}

		\subsubsection{Phase diagram}\label{sec:PhaseDiag}

From the RG flow diagram (see Fig.~\ref{fig:FlowDiag}), we calculate the phase diagram of the model in $J_zJ_x$-space, and in anisotropy-temperature space (see Fig.~\ref{fig:PhaseDiag}). In addition to the disordered paramagnetic phase (PM) at high temperatures, we identify three ordered phases at low temperatures: ferromagnetic and antiferromagnetic ordering (FM and AFM) for the Ising-like regimes of $\Delta>1$ and $\Delta<-1$, and a vortex-type ordering in the Kosterlitz-Thouless (KT) phase for the XY-like regime of $\left|\Delta\right|<1$. We emphasize that in the KT phase (also called spin-flipping phase in the presence of $U(1)$-symmetry-breaking external magnetic field \cite{Yunoki02}, or XY phase in zero-field \cite{Orrs16}), the ordering in $s^x$ (and $s^y$) components is a topological bound-vortex-antivortex ordering \cite{Kosterlitz73, Kosterlitz74} with vanishing magnetization in all directions \cite{Kar17, Cuccoli01}, but non-zero helicity modulus \cite{Roscilde03}. On the contrary, the FM (AFM) ordering results in a non-vanishing magnetization (staggered magnetization) in $s^z$ spin components \cite{Frohlich77}, as well as non-vanishing spin-spin correlations \cite{Kennedy85}. The FM and AFM phases have the usual Ising-type ground-states \cite{Manousakis91} gapped from the excited states, while the KT phase has gapless spin-liquid ground-state with algebraic power-law decay of correlations (as opposed to exponential decay) \cite{Whitlock17} and bound-vortex-antivortex pairs \cite{Betts81}. In the FM and AFM phases, the discrete $\mathbb{Z}_2$ (up$\leftrightarrow$down) symmetry is spontaneously broken, while in the KT phase it stays unbroken. The continuous $U(1)$ symmetry in the XY-like regime can be broken only by an external magnetic field (that gives rise to gapless Goldstone magnon modes), but not by decreasing temperature in zero-field. \cite{Kar17, Abouie10}

\begin{table*}[t!]
	\caption{\label{tab:Tc}Critical temperatures ($T_\text{c}=3\,(2J_x^\text{c}+|J_z^\text{c}|)^{-1}\tilde{T}$) and critical interaction parameters ($J_x^\text{c}$ or $J_z^\text{c}$) for Ising, XY, and XXX models. We observe the expected result of decreasing critical temperature with increasing spin dimensionality, \emph{i.e.}, $T_\text{c}^{(\pm\infty)}>T_\text{c}^{(0)}>T_\text{c}^{(\pm1)}$.}
	\begin{ruledtabular}
		\begin{tabular}{c c c c}
			Model & Spin dimensionality & Critical temperature & Critical interaction parameter \\
			\hline
			Ising ($|\Delta|\to\infty$, $J_x=0$) & $1$ & $T_\text{c}^{(\pm\infty)}=2.461518\,\tilde{T}$ & $J_z^\text{c}=1.218756$\\
			XY ($\Delta=0$, $J_z=0$) & $2$ & $T_\text{c}^{(0)}=0.627395\,\tilde{T}$ & $J_x^\text{c}=2.390841$\\
			XXX ($|\Delta|=1$, $J_x=|J_z|$) & $3$ & $T_\text{c}^{(\pm1)}=0$ & $J_x^\text{c}=|J_z^\text{c}|\to\infty$
		\end{tabular}
	\end{ruledtabular}
\end{table*}

\begin{table*}[t!]
	\caption{\label{tab:coeff} Fit parameters $A$ and $B$ of $T_c/\tilde{T}$ for $|\Delta|\approx1$, see text and Fig.~\ref{fig:Tc1}. Standard errors are due to fitting statistics.}
	\begin{ruledtabular}
		\begin{tabular}{c c c c c}
			Fit parameter & Ising-like AFM & XY-like AFM & XY-like FM & Ising-like FM \\
			& ($\Delta\lesssim-1$) & ($\Delta\gtrsim-1$) & ($\Delta\lesssim1$) & ($\Delta\gtrsim1$)\\
			\hline
			$A$ & $3.035\pm0.002$ & $3.1584\pm0.0008$ & $1.7432\pm0.0001$ & $1.7780\pm0.0007$\\
			$B$ & $14.55\pm0.07$ & $346.7\pm0.9$ & $36.28\pm0.02$ & $2.602\pm0.006$
		\end{tabular}
	\end{ruledtabular}
\end{table*}

Finite-temperature order-disorder transitions exist both in Ising-like regimes ($J_x<|J_z|$, including the classical Ising models $J_x=0$) and in XY-like regime ($J_x>|J_z|$, including the quantum XY model $J_z=0$). However, no finite-temperature phase transition occurs for the isotropic XXX models ($J_x=|J_z|$), for which the system becomes ordered only at zero-temperature. This fact is reflected in the flow diagram (see Fig.~\ref{fig:FlowDiag}(a)) as any point on the isotropic line with finite $J_x=J_z$, flows to the PM phase sink under successive RG transformations. Our result of the absence of finite-temperature phase transition for the two-dimensional XXX models, agrees not only with the high-temperature \cite{Yamaji73} and variational cumulant \cite{Liu94} expansion results, and effective-field RG results \cite{Sousa94}, but also with the rigorous proof by Mermin-Wagner theorem \cite{Mermin66}. We numerically confirmed that finite-temperature ordered phases for the isotropic XXX models emerge only for $d>2$, but not at $d=2$.

We can infer the special cases of Ising, XY, and XXX models as interacting spins that essentially fluctuate in one-, two-, and three-dimensions respectively. At $T=0$, the thermal fluctuations are suppressed, and all three models manifest long-range-ordered phases. This is because at $T=0$, increasing the entropy does not help minimizing the free energy, and the system is driven to ground state only by internal energy minimization. It should require less thermal energy to suppress the spin-spin orderings as the dimensionality of the spins increases. Hence, we expect the critical temperature of the Ising models ($T_\text{c}^{(\pm\infty)}$ for $|\Delta|\to\infty$) to be larger than that of the XY model ($T_\text{c}^{(0)}$ for $\Delta=0$); and the latter to be larger than the critical temperature of the XXX models ($T_\text{c}^{(\pm1)}$ for $|\Delta|=1$), which are presented in Table~\ref{tab:Tc}. The expected order $T_\text{c}^{(\pm\infty)}>T_\text{c}^{(0)}>T_\text{c}^{(\pm1)}$, is not due to the factor $3/(2+|\Delta|)$ we introduced in Hamiltonian~(\ref{eq:XXZ1}), which led to the definition~(\ref{eq:T}) of temperature. In fact, from the right-most column of Table~\ref{tab:Tc}, we see the expected order in bare interaction parameters ($J_u^\text{c}$) at the phase transitions. Furthermore, in Fig.~\ref{fig:PhaseDiag}(b), we observe that the critical temperature decreases as the absolute anisotropy parameter $|\Delta|$ is decreased from Ising models (one-dimensional spins) to XXX models (three-dimensional spins), and it decreases as $|\Delta|$ is increased from XY model (two-dimensional spins) to XXX model (three-dimensional spins). The same expected trend was also observed for square lattices by pure-quantum self-consistent harmonic approximation \cite{Cuccoli01}, by Monte Carlo simulations \cite{Ding90b}, by high-temperature series expansion \cite{Ishikawa71}, and by Pad\'{e} approximation \cite{Obokata67}, all of which give qualitatively the same anisotropy-temperature phase diagram as in Fig.~\ref{fig:PhaseDiag}.

\begin{figure}[b!]
	\includegraphics[scale=1.0]{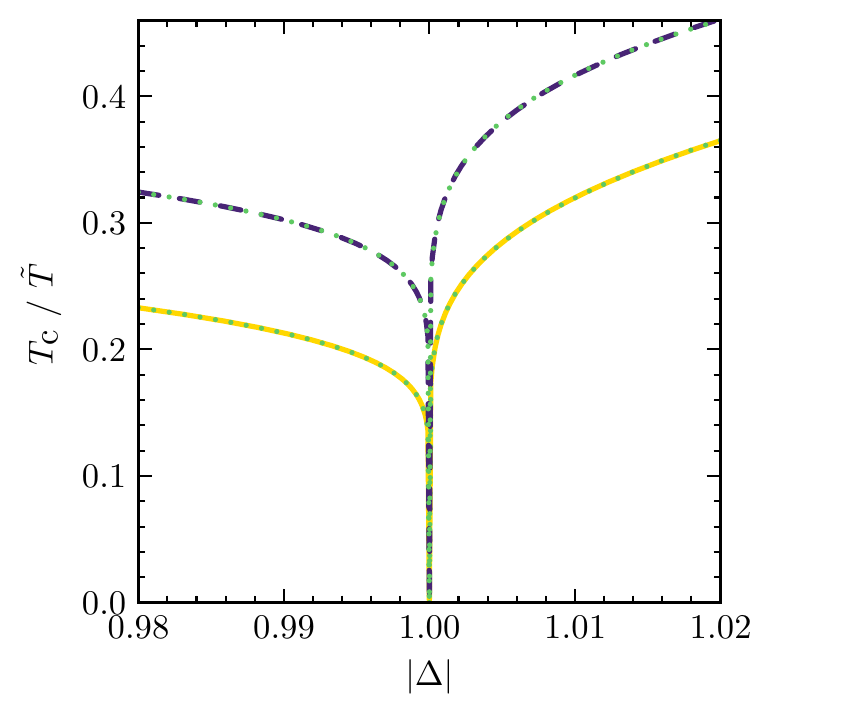}
	\caption{\label{fig:Tc1} Critical temperature $T_c/\tilde{T}$ as a function of absolute anisotropy parameter $|\Delta|$ in the vicinity of isotropic points ($|\Delta|=1$). Yellow and purple lines correspond to FM ($\Delta>0$) and AFM ($\Delta<0$) $s^z$-interactions. Overlapping dotted green lines show the fit results in the asymptotic limits $|\Delta|\to1^\pm$ (see text and Table~\ref{tab:coeff}).}
\end{figure}

It is a well-established fact that the critical temperature vanishes asymptotically as $T_\text{c}/\tilde{T}=-1/\ln(1-\Delta)$ for the \emph{classical} XXZ model ($S\to\infty$) in the limit $|\Delta|\to1$ \cite{Ding92b, Lee05, Hikami80, Kawabata82}; and similarly as $T_\text{c}/\tilde{T}=A[\ln\frac{B}{|\Delta|-1}]^{-1}$ and $T_\text{c}/\tilde{T}=A[\ln\frac{B}{1/|\Delta|-1}]^{-1}$ for the \emph{quantum} XXZ model in the limits $|\Delta|\to1^+$ and $|\Delta|\to1^-$ respectively \cite{Ding92b, Loh85a, Roscilde03}. Fitting our results to these quantum case functions in the range $0.98\leqslant|\Delta|\leqslant1.02$, yields the fit coefficients $A$ and $B$ given in Table~\ref{tab:coeff}. In Fig.~\ref{fig:Tc1}, we compare our calculated $T_\text{c}/\tilde{T}$ with fits to these functional forms, and observe a perfect agreement. We note that in the fitting range $|\Delta|=1\pm0.02$, the factor in our temperature definition $3/(2+|\Delta|)=1\pm0.007$ is very close to unity.

The logarithmic dependence of $T_\text{c}$ on $\Delta$, enables phase transitions at critical temperatures $T_\text{c}$ much away from $T=0$, even in the close vicinity of isotropic XXX models. Hence, the strong quantum fluctuations --which smear out the finite-temperature order in XXX case \cite{Mermin66, Roscilde03}-- are suppressed even by very weak anisotropy ($|\Delta|\approx1$). This finite-temperature phase transition even for very weak anisotropy is particularly important in modelling high-$T_\text{c}$ superconductors like La$_2$CuO$_4$ or monolayer quantum magnets like Ba$_3$CoSb$_2$O$_9$ (a triangular antiferromagnet) and K$_2$CuF$_4$ (a square ferromagnet), since these real systems possess very weak easy-plane or easy-axis anisotropy with $1-\left|\Delta\right|$ in the range $10^{-2}$-$10^{-5}$. \cite{Lines67, Ding90b, Ding92b, Zhou12, Susuki13}

\begin{figure}[b!]
	\includegraphics[scale=1.0]{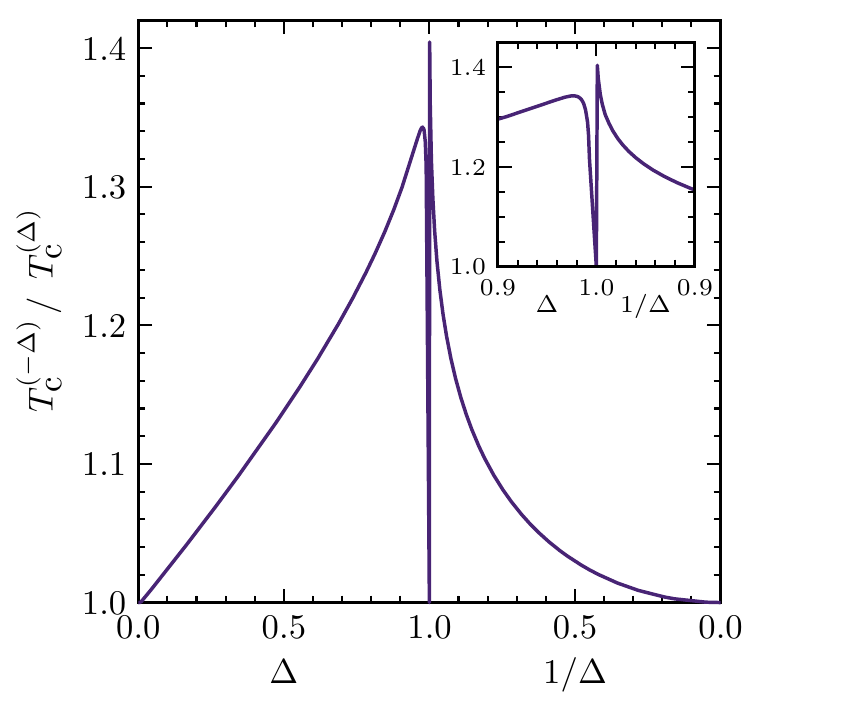}
	\caption{\label{fig:TcRatio} The ratio of AFM and FM critical temperatures, $T_\textrm{c}^{(-\Delta)}/\,T_\textrm{c}^{(\Delta)}$, as a function of positive anisotropy, $\Delta\geqslant0$. We respectively employ $\Delta$ and $1/\Delta$ as the horizontal axis in XY-like ($\Delta<1$) and Ising-like ($\Delta>1$) regimes. Inset shows the zoomed region around the isotropic XXX case, $\Delta=1$, where AFM and FM critical temperatures are both zero (see Table~\ref{tab:Tc} and Fig.~\ref{fig:Tc1}).}
\end{figure}

We note that the XY model critical temperature $T_\text{c}^{(0)}/\tilde{T}=0.627$ (see Table~\ref{tab:Tc}) we obtain for the hierarchical lattice of $b=2$ and $d=2$, is in between the values $0.591$ and $0.987$ calculated respectively for square and triangular lattices by high-temperature series expansion \cite{Rogiers79}, and also agrees with the value $0.675\pm0.075$ calculated by Monte Carlo simulations on square lattices \cite{Loh85a}. Although our result is within $6\%$-$7\%$ error range of these square lattice critical temperatures, other square lattice estimates of $0.525$ \cite{Ding90a} and $0.529$ \cite{Ding92a} by Monte Carlo simulations, and $0.54$ by pure-quantum self-consistent harmonic approximation \cite{Cuccoli95} are about $20\%$ error range to our $T_\text{c}^{(0)}/\tilde{T}$ result. We recall that the Suzuki-Takano approach works at its worst for $\Delta=0$. In the opposite limit, the critical temperature $T_\text{c}^{(\pm\infty)}/\tilde{T}=2.46$ obtained for the Ising models, is $8\%$ close to Onsager's exact value $2/\ln(1+\sqrt{2})\approx2.27$ for square lattice \cite{Onsager44}. 

We also note that although the critical temperatures for FM ($\Delta\to\infty$) and AFM ($\Delta\to-\infty$) Ising models are the same, this symmetry disappears for finite $\Delta$ introduced by non-zero $J_x$. That said, phase diagrams shown in Fig.~\ref{fig:PhaseDiag} do not possess mirror symmetries about $J_z=0$ and $\Delta=0$ lines (also see Fig.~\ref{fig:Tc1} for the comparison of $T_\text{c}^{(\pm\Delta)}$ at $|\Delta|\approx1$). In example, the critical temperatures are $T_\text{c}^{(0.5)}=0.451301\,\tilde{T}$ and $T_\text{c}^{(2)}=1.124462\,\tilde{T}$ for the FM $\Delta=0.5$ and $\Delta=2$, while they are $T_\text{c}^{(-0.5)}=0.515006\,\tilde{T}$ and $T_\text{c}^{(-2)}=1.158044\,\tilde{T}$ for the AFM $\Delta=-0.5$ and $\Delta=-2$. In general, we observe that the critical temperature $T_\text{c}^{(\Delta)}$ for a positive finite anisotropy $\Delta$ is always smaller than $T_\text{c}^{(-\Delta)}$ (except for $\Delta=1$, for which $T_\text{c}^{(\pm1)}=0$, see Table~\ref{tab:Tc}). This asymmetry between FM and AFM critical temperatures is a pure quantum mechanical effect, and was indeed signaled by the specific heat peaks in $d=1$: specific heat peak temperatures shift in opposite directions for AFM and FM models as the anisotropy gets stronger in $d=1$. \cite{Sariyer08}

As will be demonstrated in Section~\ref{sec:ZeroT}, the quantum fluctuations at $T=0$ are stronger in the AFM case compared to the FM case, as expected \cite{Frohlich77, Vogt09}. Hence, one expects the AFM order at low-temperatures to be washed out more easily (less thermal energy $kT_\text{c}$ is required) compared to FM order. Thus, one expects $T_\text{c}^{(-\Delta)}/\,T_\text{c}^{(\Delta)}\leqslant1$. However, we observe the opposite case: $T_\text{c}^{(-\Delta)}/\,T_\text{c}^{(\Delta)}\geqslant1$. This well-known result might therefore cannot be explained by the strength of quantum fluctuations. A possible explanation might be due to interrelated effects of entropy and density of states. \cite{Vogt09} In Fig.~\ref{fig:TcRatio}, we plot the ratio of AFM and FM critical temperatures, $T_\text{c}^{(-\Delta)}/\,T_\text{c}^{(\Delta)}$, as a function of anisotropy parameter $\Delta$. We see that this ratio increases from unity as $\Delta$ increases from zero, makes a dip at the XXX point ($\Delta=1$), and further decreases back to unity as Ising models are approached ($\Delta\to\infty$), where there are no quantum effects. We expect the dip at $\Delta=1$ to be smoothed out at higher dimensions ($d>2$), where an order-disorder phase transition occurs for XXX models ($\Delta=\pm1$). As a comparison to data in Fig.~\ref{fig:TcRatio}, for $d=3$, the ratio $T_\text{c}^{(-1)}/\,T_\text{c}^{(1)}$ was estimated to be $1.48$ \cite{Kaplan08}, $1.22$ \cite{Hinczewski05, Falicov95}, $1.13$ \cite{Oitmaa04}, $1.12$ \cite{Rushbrooke63, Charles72, Kasteleijn56a, *Kasteleijn56b}, and $1.11$ \cite{Mano90}.

\begin{figure*}[t!]
	\begin{tabular}{c}
		\includegraphics[scale=1.0]{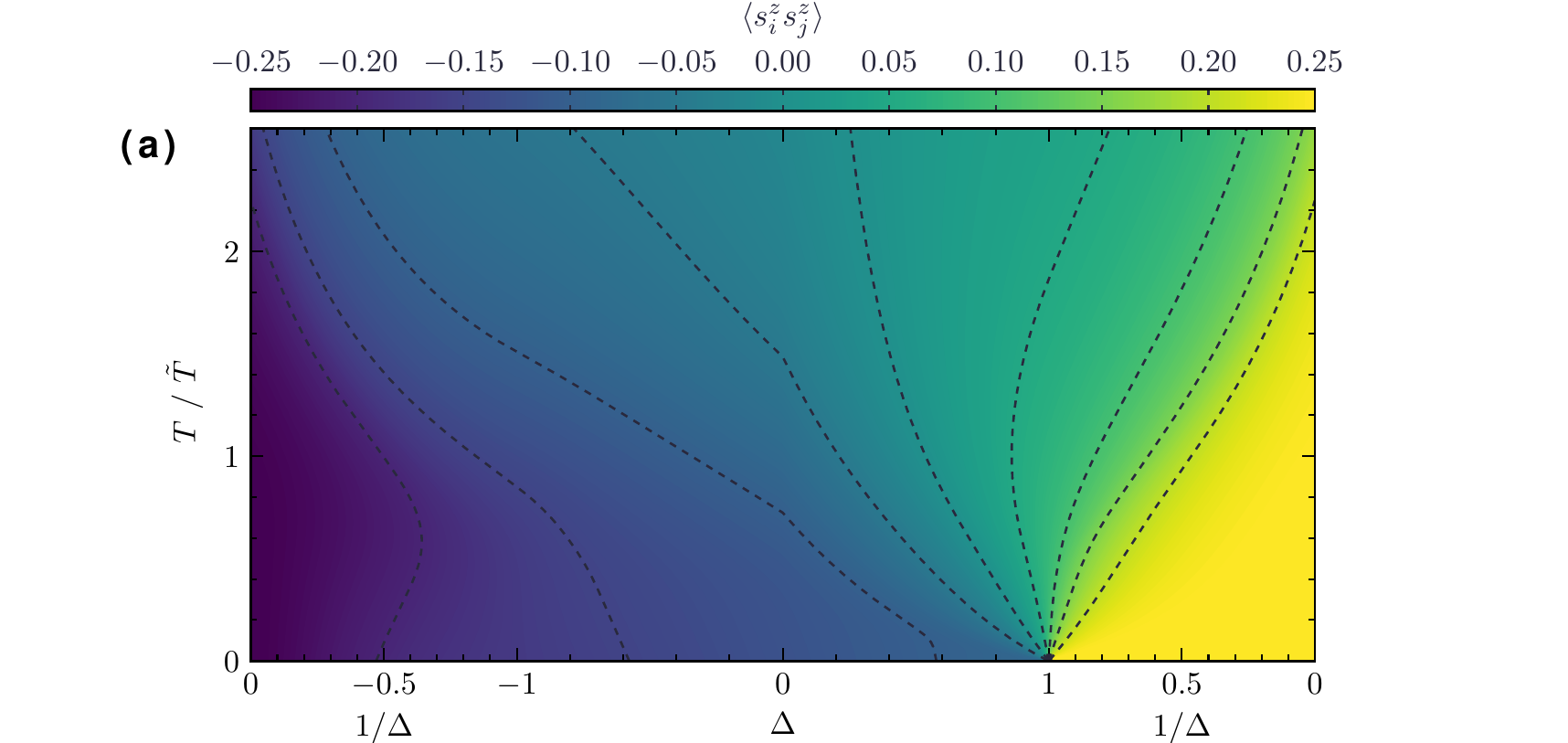}\\
		\\
		\includegraphics[scale=1.0]{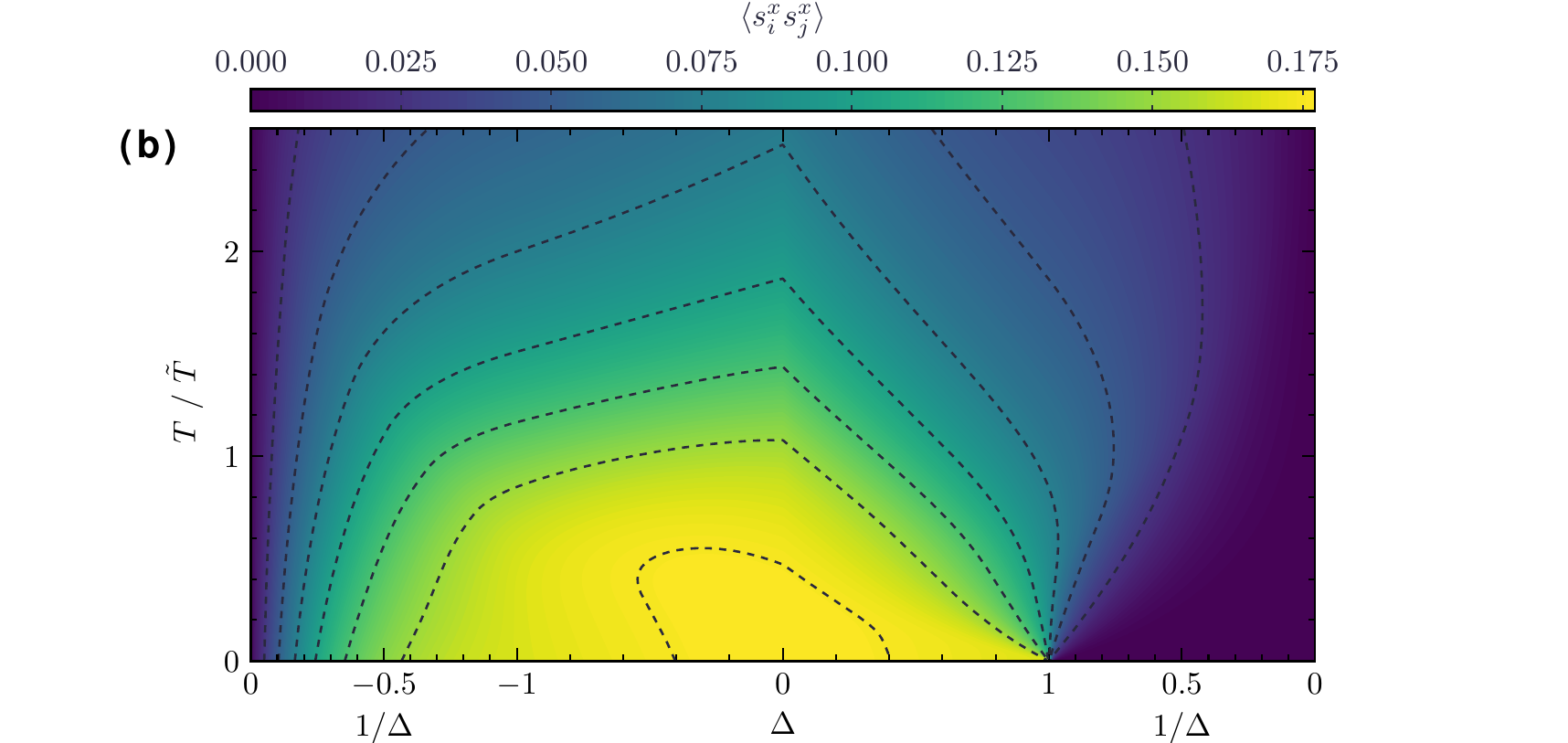}\\
	\end{tabular}
	\caption{\label{fig:CorrCont} Nearest-neighbor spin-spin correlations $\langle s_i^zs_j^z\rangle$ (a) and $\langle s_i^xs_j^x\rangle$ (b) as functions of anisotropy parameter and temperature. We respectively employ $\Delta$ and $1/\Delta$ as the horizontal axes in XY-like ($|\Delta|<1$) and Ising-like ($|\Delta|>1$) regimes. For both correlations, note the cusps at $\Delta=0$ and the singularities at $\Delta=1$, $T=0$.}
\end{figure*}

\begin{figure*}[t!]
	\includegraphics[scale=1.0]{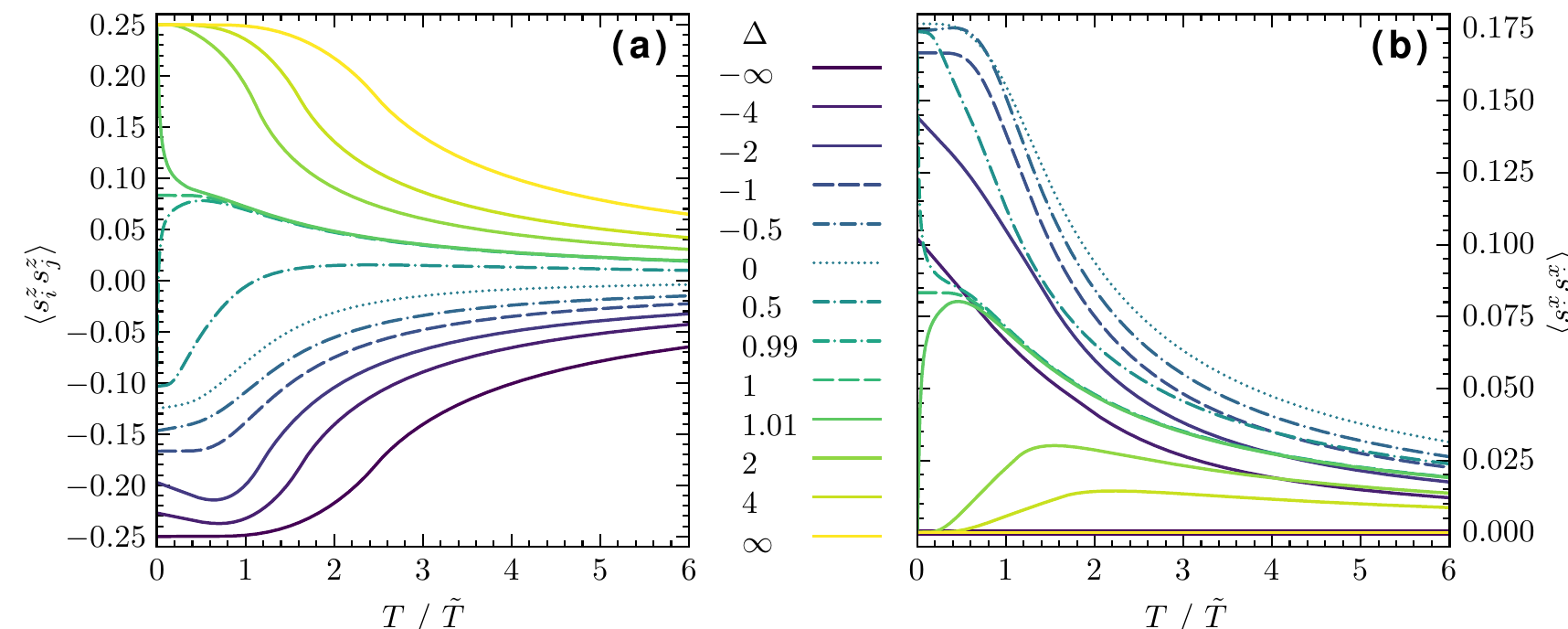}
	\caption{\label{fig:CorrAll} Nearest-neighbor spin-spin correlations $\langle s_i^zs_j^z\rangle$ (a) and $\langle s_i^xs_j^x\rangle$ (b) as functions of temperature for anisotropy parameters $\Delta$ indicated in the legend. Note that $\langle s_i^xs_j^x\rangle=0$ at all temperatures, in both AFM and FM Ising limits ($\Delta\to\pm\infty$). Also note the discontinuities in both correlations at $\Delta=1$, $T=0$.}
\end{figure*}

	\subsection{Thermodynamic functions}
		\subsubsection{Spin-spin correlations} \label{sec:Corr}
Using equation~(\ref{eq:original}), we calculated the nearest-neighbor spin-spin correlations, $\langle s_i^zs_j^z\rangle$ and $\langle s_i^xs_j^x\rangle$. Results are presented in Fig.~\ref{fig:CorrCont} as contour plots. At zero-temperature, we observe a discontinuity in both correlations at $\Delta=1$, as the anisotropy passes from XY-like regime ($|\Delta|<1$) to FM Ising-like regime ($\Delta>1$). This discontinuity has also been observed before for one-dimensional XXZ model \cite{Sariyer08}, and will be discussed in detail in Section~\ref{sec:ZeroT}.

In Fig.~\ref{fig:CorrAll}, we plot nearest-neighbor spin-spin correlations $\langle s_i^zs_j^z\rangle$ and $\langle s_i^xs_j^x\rangle$ as functions of temperature, for various anisotropies spanning the whole Ising-like and XY-like regimes. We observe positive $\langle s_i^xs_j^x\rangle>0$ globally, except in the (A)FM Ising limits ($J_x=0$), where $\langle s_i^xs_j^x\rangle$ vanishes for all temperatures as expected. As $\Delta$ increases from the AFM Ising limit ($\Delta\to-\infty$) to the XY model ($\Delta=0$), $\langle s_i^xs_j^x\rangle$ correlations grow, which is due to increasing $J_x$ compared to $\left|J_z\right|$. As $\Delta$ further increases from the XY model ($\Delta=0$) to the FM Ising limit ($\Delta\to\infty$), $\langle s_i^xs_j^x\rangle$ correlations decline back to zero, which is due to decreasing $J_x$ compared to $J_z$.

In the XY model ($J_z=0$), the $\langle s_i^xs_j^x\rangle$ correlations max out to $\langle s_i^xs_j^x\rangle=0.176777$ at zero-temperature (which compares well with the value 0.141 obtained for a 16-site square lattice \cite{Oitmaa78}). This value is $1/\sqrt{2}$ ($70\%$) times the possible maximum $\langle s_i^us_j^u\rangle=1/4$ for spin-$\nicefrac{1}{2}$ correlations. For isotropic interactions in two spin dimensionalities ($x$ and $y$), the correlation $\langle s_i^xs_j^x\rangle$ cannot attain this paramount value in the algebraically ordered KT phase, which is characterized by vanishing magnetization with emerging vortex-antivortex pairs. We note that the paramount correlations value $1/4$ can be reached only at zero-temperature by $\langle s_i^zs_j^z\rangle$ correlations in FM Ising-like regime, $\Delta>1$. Recall that in the extreme limit $\Delta\to\pm\infty$, the quantum XXZ model is reduced to classical Ising model, \emph{i.e.}, interactions of one-dimensional classical spins. Similarly, $\langle s_i^zs_j^z\rangle$ become $-1/4$ at $T=0$ in the AFM Ising limit $\Delta\to-\infty$ (see Fig.~\ref{fig:CorrAll}(a)).

The correlations $\langle s_i^zs_j^z\rangle$ increase as $\Delta$ increases from AFM Ising limit to FM Ising limit. In the ferromagnetic XY-like regime $0\leqslant\Delta<1$, although both interactions $J_x>0$ and $J_z\geqslant0$ are non-negative, the $\langle s_i^zs_j^z\rangle$ correlations are negative at low temperatures. This fact is a consequence of quantum fluctuations dominating over thermal fluctuations at low temperatures, and  $\langle s_i^zs_j^z\rangle$ correlations become positive at higher temperatures (see Fig.~\ref{fig:CorrAll}(a)), where thermal fluctuations get stronger.

\begin{figure*}[t!]
	\includegraphics[scale=1.0]{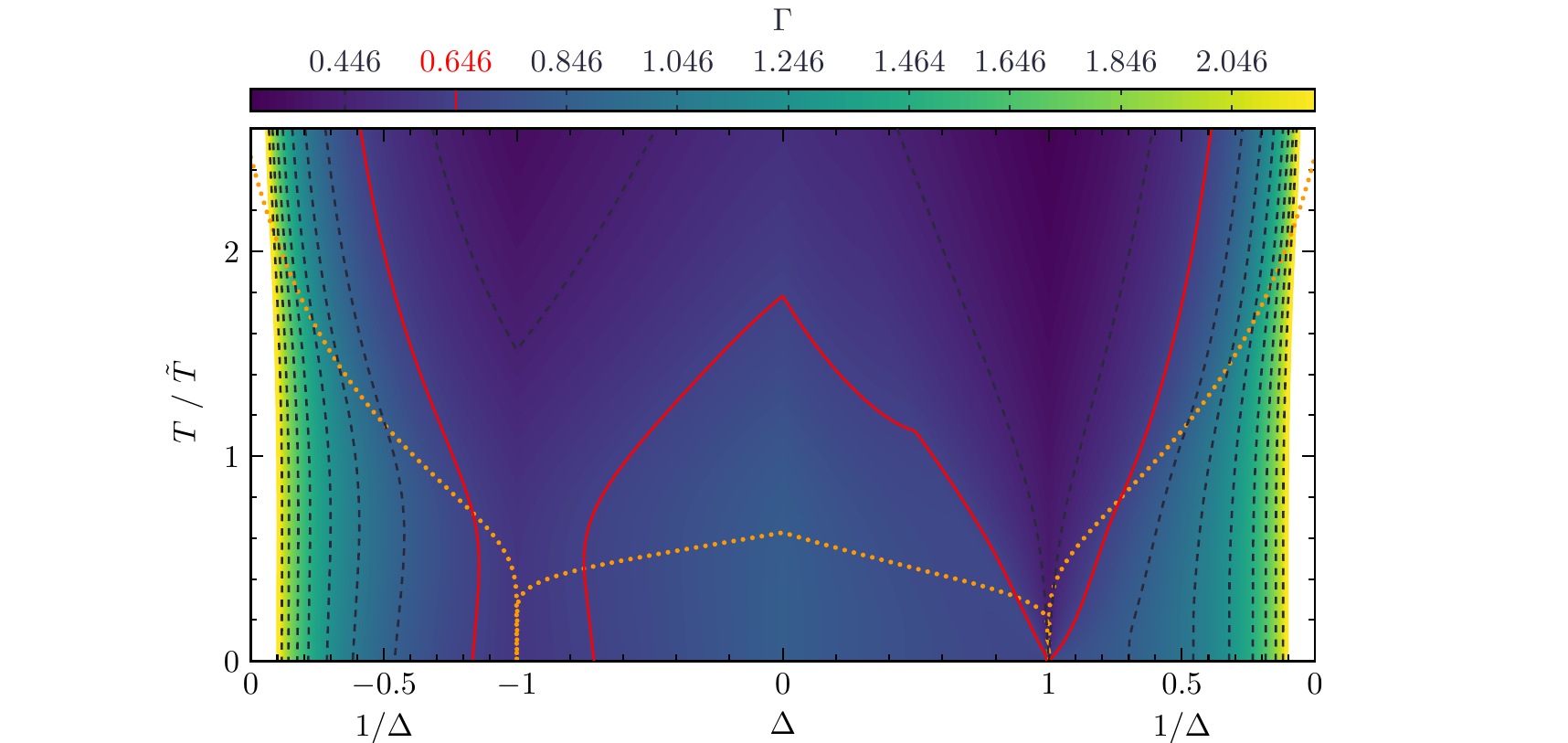}
	\caption{\label{fig:lro} Long-range-order measure $\Gamma$ as a function of anisotropy and temperature. We employ $\Delta$ and $1/\Delta$ as the horizontal axis in XY- ($|\Delta|<1$) and Ising-like ($|\Delta|>1$) regimes respectively. Solid red contours mark the approximate lower bound for long-range-order threshold, $\Gamma_2\approx0.646$. Phase boundaries (see Fig.~\ref{fig:PhaseDiag}(b)) are superposed as the dotted orange line.}
\end{figure*}

One particularly interesting result is the non-vanishing $\langle s_i^zs_j^z\rangle<0$ correlations at all temperatures for $\Delta=0$ (XY model). In the classical XY (planar) model, spins are restricted to fluctuate only in the $xy$-plane. In contrast, in the quantum XY model, commutation between the out-of-plane $s^z$ and in-plane $s^{x(y)}$ operators give rise to quantum fluctuations in $s^z$-components. As a consequence of this pure quantum mechanical effect, critical temperature for quantum XY model ($T_\text{c}^{(0)}/\tilde{T}=0.627$, see Table~\ref{tab:Tc}) is lower than that of the classical planar model ($1.796$ \cite{Ding90a, Ding92a}); and we have non-vanishing $\langle s_i^zs_j^z\rangle$ correlations at all finite-temperatures, even for the XY model with no interactions between $s^z$ components ($\Delta=0$, $J_z=0$). For any value of the anisotropy parameter $\Delta$, the finite $\langle s_i^zs_j^z\rangle$ correlations at low temperatures decay --as expected-- as the temperature is increased. Such a decay also applies for the $\langle s_i^xs_j^x\rangle$ correlations.

Several thermodynamic and entanglement functions can be deduced from the spin-spin correlations. Below, we will discuss the global finite-temperature behavior of long-range-order and entanglement measures, internal energy density, and specific heat, before moving to low-temperature and ground-state properties.

		\subsubsection{Long-range-order measure}
Long-range-order can be identified using the measure
\begin{equation}
\label{eq:lro}
\Gamma=\begin{cases}
\sqrt{\left|2\,\Delta\,\langle s_i^zs_j^z\rangle\right|} & , \left|\Delta\right|\geqslant1 \\
\frac{2\,\langle s_i^xs_j^x\rangle}{\sqrt{\langle s_i^xs_j^x\rangle+\left|\Delta\,\langle s_i^zs_j^z\rangle\right|}} & , \left|\Delta\right|\leqslant1
\end{cases}
\end{equation}
in the Ising-like ($\left|\Delta\right|>1$) and XY-like ($\left|\Delta\right|<1$) regimes \cite{Kubo88, Ozeki89, Nishimori89, Wischmann91}, although it is known that the two-dimensional XY model ($\Delta=0$) at low-temperatures has topological KT-type order rather than long-range order \cite{Mermin66, Kosterlitz73, Kosterlitz74, Cuccoli01}. Nevertheless, it was rigorously proven by Kennedy, Lieb and Shastry that the XY model has ground-state long-range-order at \emph{zero-temperature} for $d>1$. \cite{Kennedy88} Hence, if $\Gamma$ is larger than a threshold $h_d$ in the Ising-like regime, then long-range-order exists in $\langle s_i^zs_j^z\rangle$; and we can infer that if $\Gamma>h_d$ in the XY-like regime, then \emph{quasi}-long-range-order exists in $\langle s_i^xs_j^x\rangle$ correlations \cite{Takada80}. In $d=2$ dimensions, a lower bound for $h_2$ is estimated as $\Gamma_2\approx0.646$. \cite{Kubo88, Ozeki89, Nishimori89, Wischmann91}

In Fig.~\ref{fig:lro}, we plot long-range-order measure~(\ref{eq:lro}), calculated using the spin-spin correlation results of Section~\ref{sec:Corr}, as a function of anisotropy and temperature. We clipped the plot at $\Gamma=2.2$, since as Ising models are approached ($1/\Delta\to0^\pm$), $\Gamma$ grows very fast, \emph{e.g.}, becomes $\Gamma\approx7$ for $1/\Delta=\pm0.01$ at low temperatures.

In Fig.~\ref{fig:lro}, we observe that at low enough temperatures, $\Gamma>\Gamma_2$ for all $\Delta$, except in the vicinity of the AFM XXX point ($\Delta=-1$), where the approximation of $h_2$ by $\Gamma_2$ must be broken down. Hence, we conclude qualitatively that Ising-like (XY-like) ordered phases AFM and FM (KT) in Fig.~\ref{fig:PhaseDiag}, manifest log-range (quasi-long-range) order in $\langle s_i^zs_j^z\rangle$ ($\langle s_i^xs_j^x\rangle$) correlations.

Although the order-disorder transition temperatures $T_\text{c}^{(\Delta)}$ can in principle be calculated by the identity $\Gamma(\Delta,T_\text{c}^{(\Delta)})=h_2$, such a calculation requires the exact $h_2=h_2(\Delta)$, which is actually a function of anisotropy parameter. However, even from the approximate relation $\Gamma(\Delta,T_\text{c}^{(\Delta)})=\Gamma_2$ (see red contour lines in Fig.~\ref{fig:lro}), we observe two expected results: (\emph{i}) that the critical temperature (between disordered and long-range-ordered phases) monotonically decreases as $\Delta$ approaches to $\pm1$ both from Ising-like and XY-like regimes, and (\emph{ii}) that the critical temperature for Ising models is higher than that for the XY model.

In Fig.~\ref{fig:lro}, for a comparison to $\Gamma=\Gamma_2$ contours (solid red lines), we also superimpose the order-disorder phase boundaries (dotted orange lines, see Fig.~\ref{fig:PhaseDiag}(b)). We see that $T_\text{c}^{(\Delta)}$ obtained approximately from $\Gamma(\Delta,T_\text{c}^{(\Delta)})=\Gamma_2$ is an underestimation (overestimation) in the vicinity of (away from) XXX models. Values for the threshold $h_2(\Delta)$ can be found from the $\Gamma$ values on the phase boundaries, \emph{i.e.}, by the identity $h_2(\Delta)=\Gamma(\Delta,T_\text{c}^{(\Delta)})$. From Fig.~\ref{fig:lro}, we see that the threshold changes non-monotonically in the interval $0.4\lesssim h_2(\Delta)\lesssim2$ for $-10<\Delta<10$, but we should emphasize that $h_2(\Delta)$ diverges in Ising limits.

		\subsubsection{Nearest-neighbor quantum entanglement measures}

Using the spin-spin correlation results (see Section~\ref{sec:Corr}), we calculated the entanglement measures for nearest-neighboring spins, which are useful for quantum computational applications \cite{Amico08, Roscilde04}. These are the entanglement of formation $\mathcal{E}$ and the quantum discord $\mathcal{D}$.

As defined by Wootters \cite{Wootters98}, $\mathcal{E}$ is the minimum average entanglement of an ensemble of pure states that represents a mixed state, which can be calculated as \cite{Fanchini10, Werlang10}
\begin{equation}
\mathcal{E}=-g\left[f\left(\mathcal{C}\right)\right]-g\left[1-f\left(\mathcal{C}\right)\right],
\end{equation}
where $g\left[f\right]=f \log_2f$ and $f\left(\mathcal{C}\right)=\tfrac{1}{2}\left(1+\sqrt{1-\mathcal{C}^2}\right)$. Here, the concurrence $\mathcal{C}$ is an entanglement monotone defined as \cite{Werlang10, Justino12} $\mathcal{C}=\max\left[0 ,4\left|\langle s_i^xs_j^x\rangle\right|-\tfrac{1}{2}\left|1+4\langle s_i^zs_j^z\rangle\right|\right]$, which can be taken as a measure of entanglement itself \cite{Wootters98}. We numerically checked for the XXZ model that $\mathcal{C}$ and $\mathcal{E}$ have the same qualitative behavior globally, \emph{i.e.}, they increase and decrease together, and hence they have the same extrema and discontinuities.

\begin{figure*}[t!]
	\begin{tabular}{c}
		\includegraphics[scale=1.0]{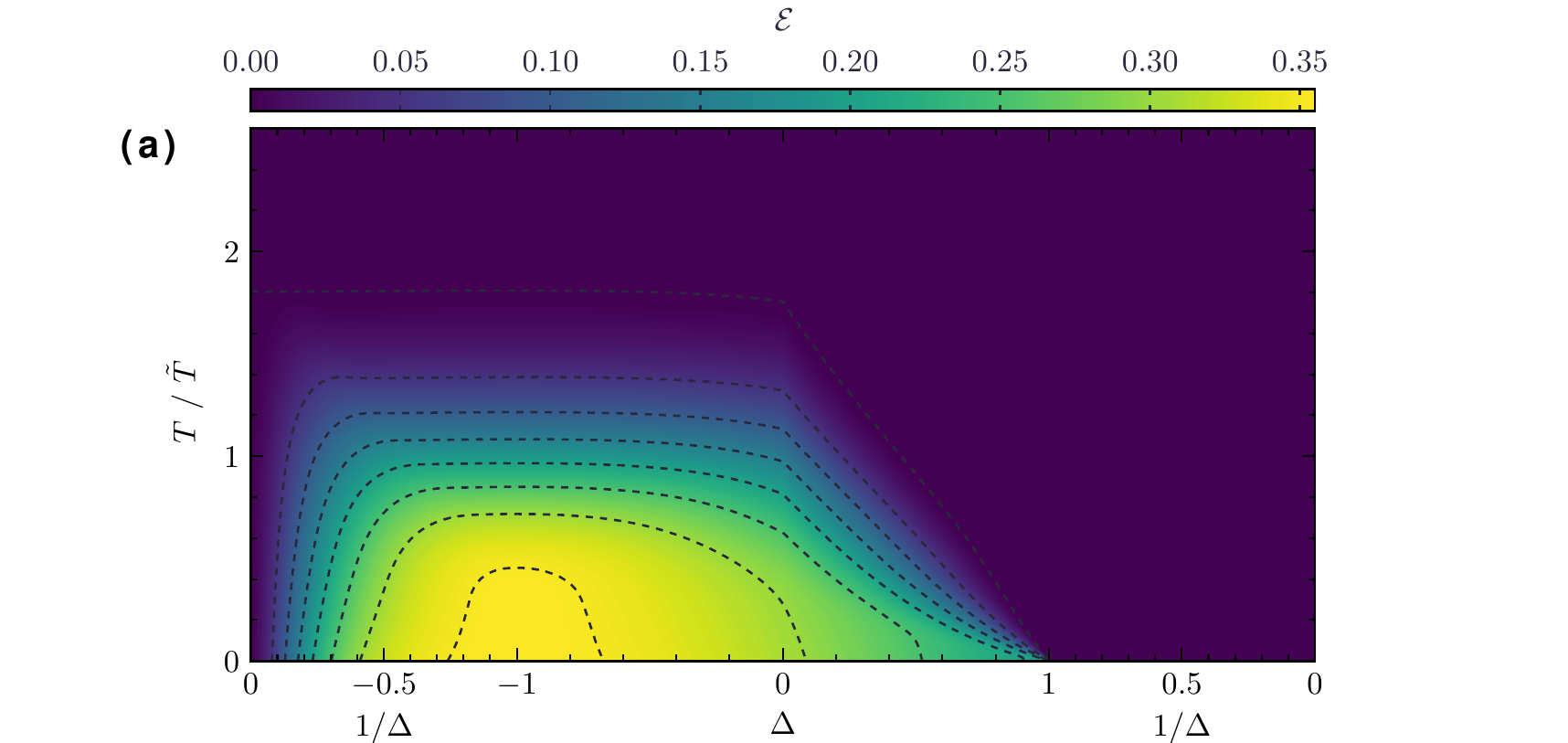}\\
		\\
		\includegraphics[scale=1.0]{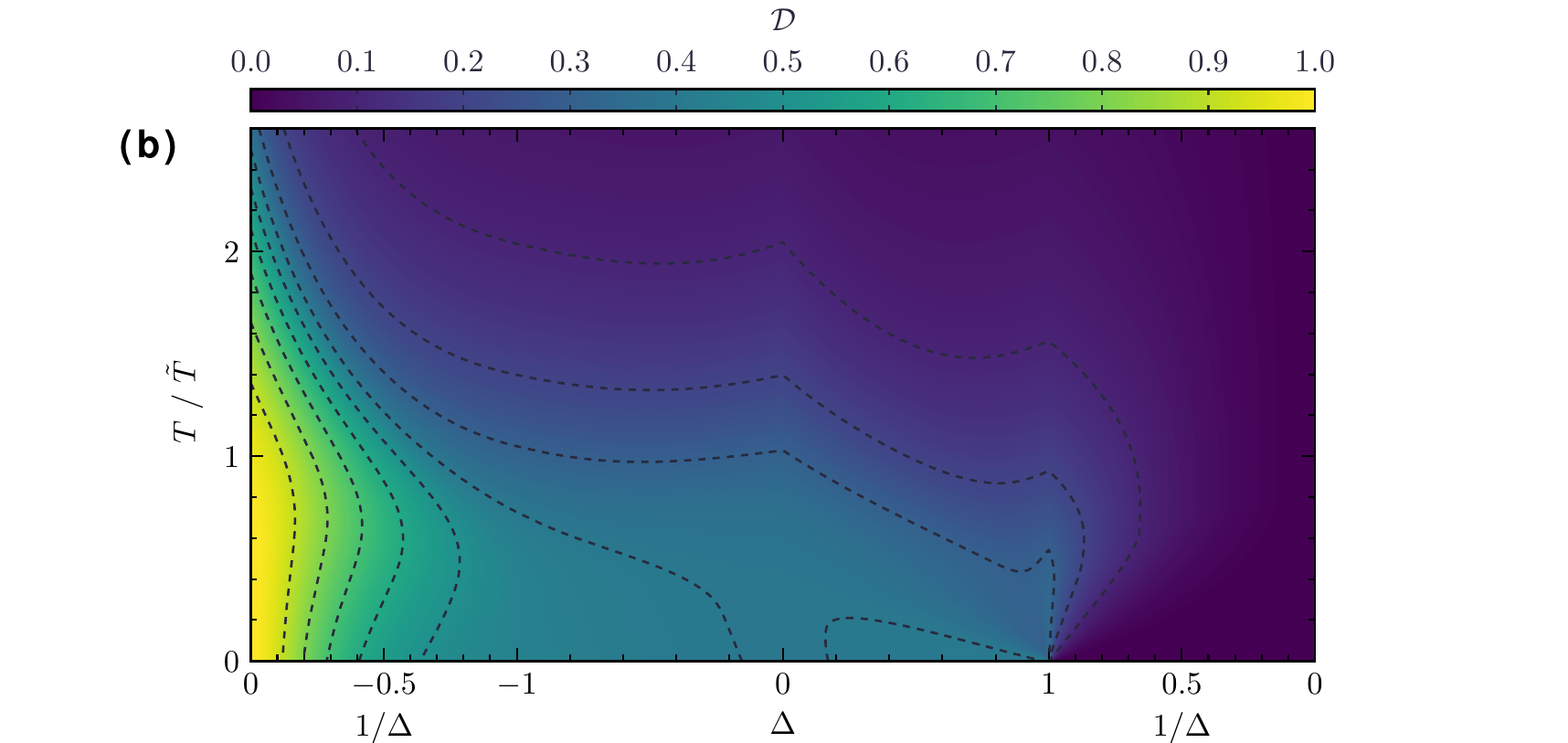}\\
	\end{tabular}
	\caption{\label{fig:QEntCont} Entanglement of formation $\mathcal{E}$ (a) and quantum discord $\mathcal{D}$ (b) as functions of anisotropy and temperature. We respectively employ $\Delta$ and $1/\Delta$ as the horizontal axes in XY-like ($|\Delta|<1$) and Ising-like ($|\Delta|>1$) regimes. Singularities and cusps in spin-spin correlations (see Fig.~\ref{fig:CorrCont}) are present in entanglement measures as well. Note the additional finite-temperature cusp in $\mathcal{D}$ at $\Delta=1$.}
\end{figure*}

\begin{figure*}[t!]
	\includegraphics[scale=1.0]{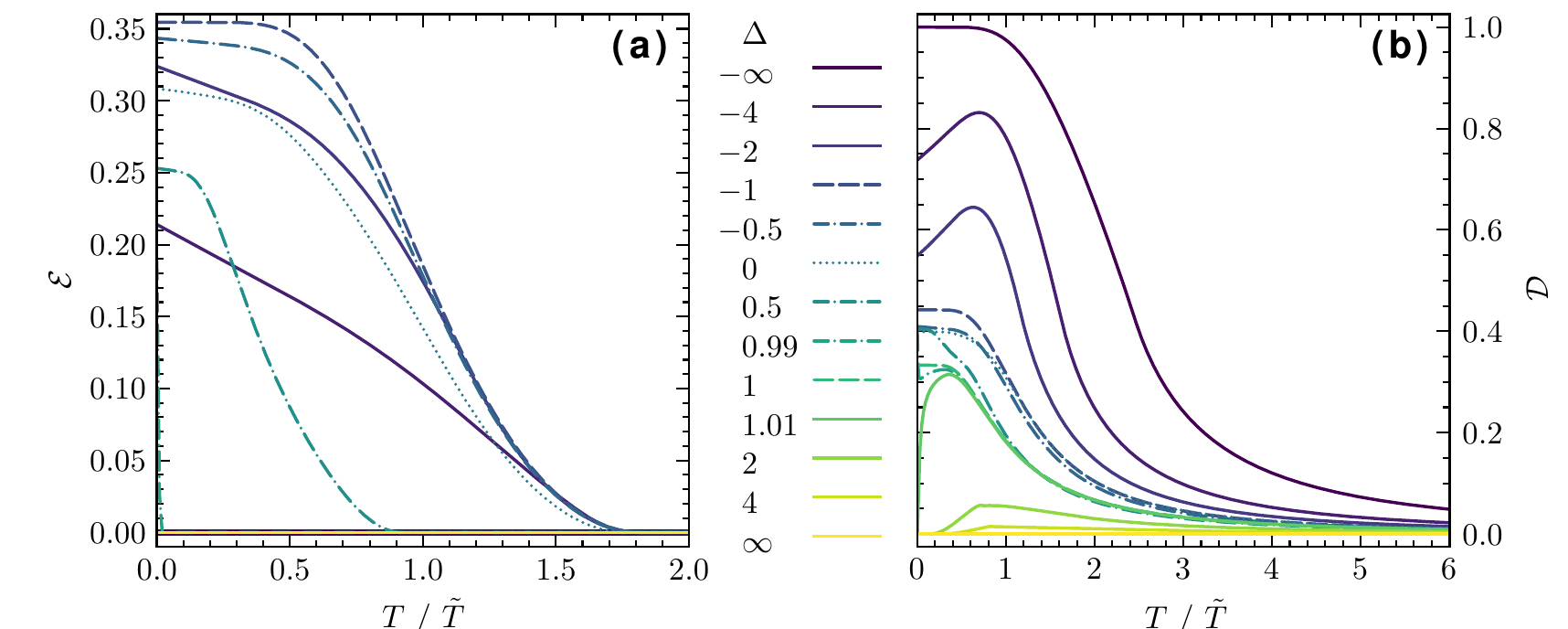}
	\caption{\label{fig:QEntAll} Entanglement of formation $\mathcal{E}$ (a) and quantum discord $\mathcal{D}$ (b) as functions of temperature for anisotropy parameters $\Delta$ indicated in the legend. Note that $\mathcal{E}=0$ at all temperatures for AFM Ising model ($\Delta\to-\infty$) and for FM Ising-like regime ($\Delta\geqslant1$), while $\mathcal{D}=0$ at all temperatures only for FM Ising model ($\Delta\to\infty$).}
\end{figure*}

Introduced by Ollivier and Zurek \cite{Ollivier01}, another measure of quantum correlations is the quantum discord $\mathcal{D}$, which measures quantum correlations due to non-commutativity, instead of entanglement \cite{Luo08}. It measures the contrast between two distinct quantum analogs of the classical mutual information, and it can be calculated from the spin-spin correlations as \cite{Werlang10, Luo08}
\begin{multline}
\mathcal{D}=\frac{1}{4}\Big[
g\left(1-8\langle s_i^xs_j^x\rangle-4\langle s_i^zs_j^z\rangle\right)+2g\left(1+4\langle s_i^zs_j^z\rangle\right)\\
+g\left(1+8\langle s_i^xs_j^x\rangle-4\langle s_i^zs_j^z\rangle\right)-2\left[g_++g_-\right]\Big],
\end{multline}
where $g_\pm=g\left(1\pm4\max\left[|\langle s_i^xs_j^x\rangle|, |\langle s_i^zs_j^z\rangle|\right]\right)$. Results for $\mathcal{E}$ and $\mathcal{D}$ are shown in Fig.~\ref{fig:QEntCont} as contour plots. In Fig.~\ref{fig:QEntAll}, we plot them as functions of temperature, for various $\Delta$ spanning the whole Ising-like and XY-like regimes.

We observe vanishing $\mathcal{E}$ in the Ising limits ($\Delta\to\pm\infty$) for all temperatures. We also observe that $\mathcal{E}$ vanishes at high temperatures ($T\gtrsim1.8\,\tilde{T}$) for all anisotropies. This is an expected result due to quantum fluctuations being overwhelmed by thermal fluctuations at high temperatures. An interesting result is vanishing $\mathcal{E}$ for the whole FM Ising-like regime ($\Delta\geqslant1$) at all temperatures, even for finite $\Delta$ at $T=0$. In the $\Delta>1$ regime at $T=0$, we have constant $\langle s_i^xs_j^x\rangle=$0 and $\langle s_i^zs_j^z\rangle=1/4$ (see Figs. \ref{fig:CorrCont}, \ref{fig:CorrAll}, and \ref{fig:ZeroT}), which translates into zero-concurrence, and thus into $\mathcal{E}=0$. For other regimes ($\Delta<1$), $\mathcal{E}$ increases with decreasing temperature, and becomes maximal at $T=0$, where thermal fluctuations vanish, and the spin-spin correlations are of pure quantum mechanical nature.

On the other hand, $\mathcal{D}$ vanishes at all temperatures only for FM Ising model ($\Delta\to\infty$). It increases with decreasing $\Delta$ in the FM and AFM Ising-like regimes ($1<\Delta<\infty$ and $-\infty<\Delta<-1$), while it changes non-monotonically in the XY-like regime ($-1<\Delta<1$). It attains the maximum possible value $\mathcal{D}=1$ for the AFM Ising model at $T=0$. In Fig.~\ref{fig:QEntAll}, we see that $\mathcal{D}$ decays with increasing $T$ as expected. Like $\mathcal{E}$, $\mathcal{D}$ also vanishes at $T=0$ for $\Delta>1$, which is due to aforementioned $\langle s_i^xs_j^x\rangle=$0 and $\langle s_i^zs_j^z\rangle=1/4$ in this region (see Section~\ref{sec:ZeroT}). However, unlike $\mathcal{E}$, $\mathcal{D}$ increases first, and then declines back as the temperature is increased. This non-monotonic behavior in Ising-like FM regime ($\Delta>1$) has also been obtained for two-qubit systems, for which, increasing $\mathcal{D}$ with decreasing $\Delta$ was also observed \cite{Werlang10}, as we do in Fig.~\ref{fig:QEntAll}.

		\subsubsection{Internal energy} \label{sec:Uint}

\begin{figure*}[t!]
	\begin{tabular}{c}
		\includegraphics[scale=1.0]{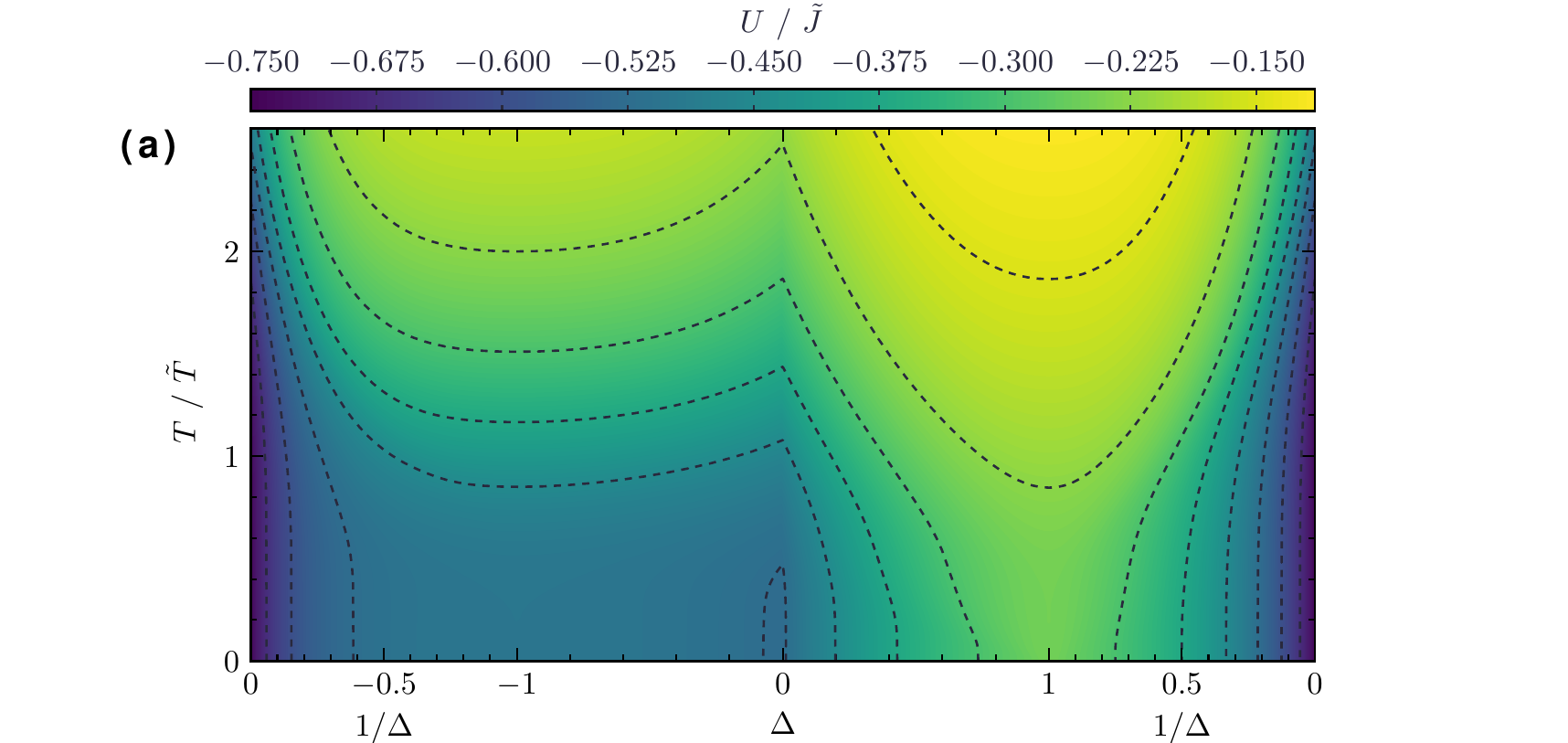}\\
		\\
		\includegraphics[scale=1.0]{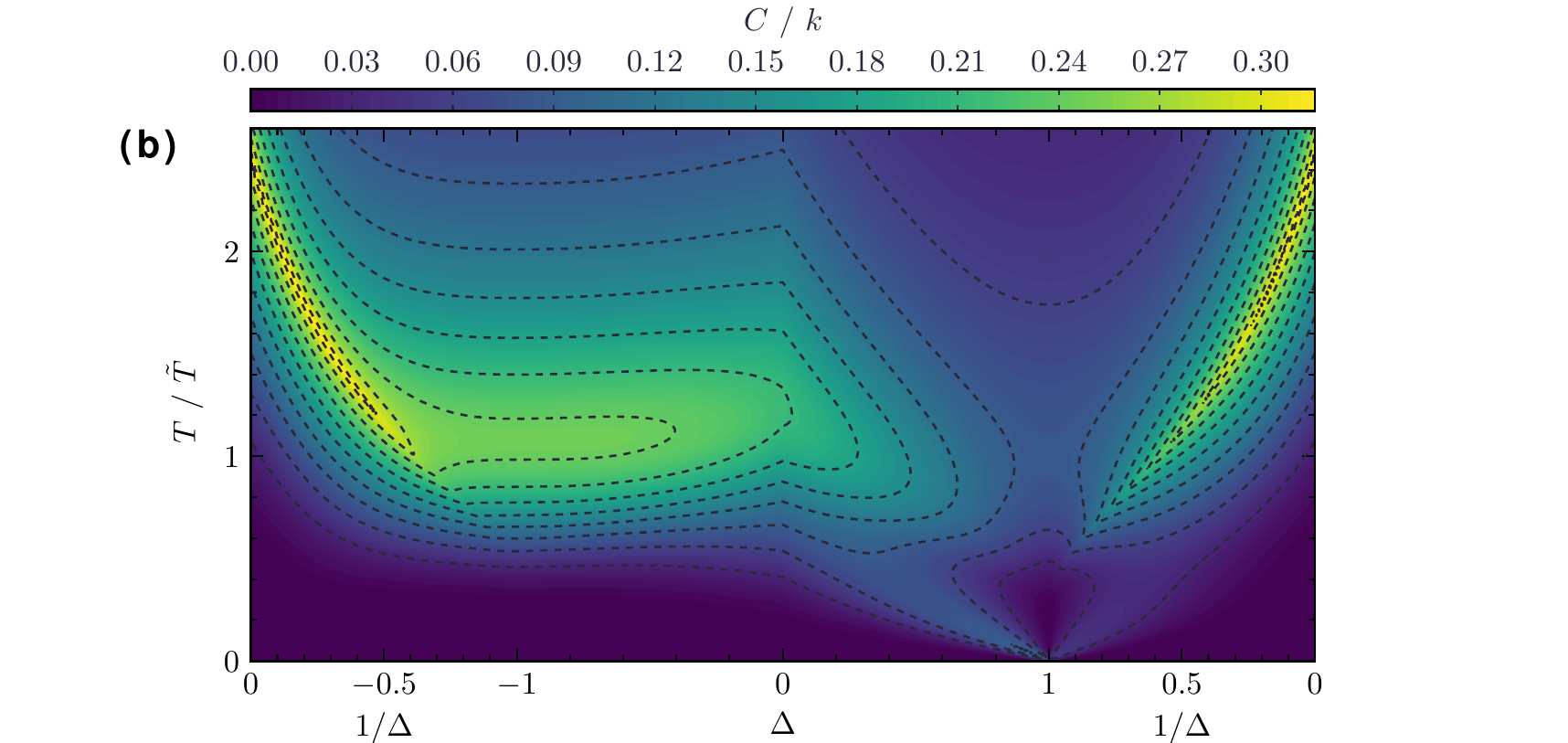}\\
	\end{tabular}
	\caption{\label{fig:UiCsCont} Internal energy density $U$ scaled by the material-dependent nearest-neighbor exchange interaction energy $\tilde{J}$ (a) and specific heat $C$ scaled by Boltzmann constant $k$ (b) as functions of anisotropy and temperature. We respectively employ $\Delta$ and $1/\Delta$ as the horizontal axes in XY-like ($|\Delta|<1$) and Ising-like ($|\Delta|>1$) regimes. The discontinuity in correlations (see Fig.~\ref{fig:CorrCont}) translates into cusps along the $\Delta$-axis at $\Delta=1$, $T=0$. The cusps at $\Delta=0$ are due to cusps in correlations (see Fig.~\ref{fig:CorrCont}).}
\end{figure*}

\begin{figure*}[t!]
	\includegraphics[scale=1.0]{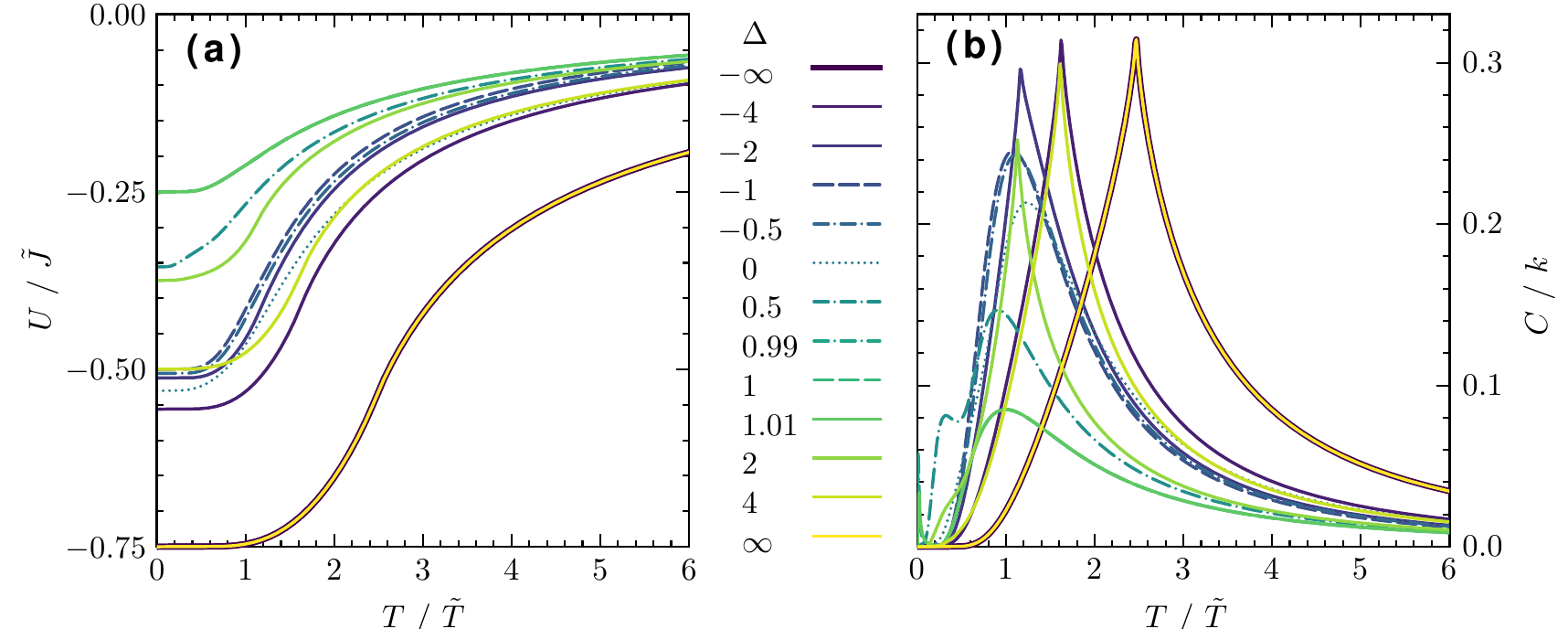}
	\caption{\label{fig:UiCsAll} Internal energy $U$ per bond scaled by the material-dependent nearest-neighbor exchange interaction energy $\tilde{J}$ (a) and specific heat $C$ scaled by Boltzmann constant $k$ (b) as functions of temperature for anisotropy parameters $\Delta$ indicated in the legend. Note that $U/\tilde{J}$ (and thus $C/k$) for AFM and FM Ising models ($\Delta\to\pm\infty$) are the same.}
\end{figure*}

We calculated the dimensionless internal energy density (per bond) $U/\tilde{J}$ from the spin-spin correlations via equation~(\ref{eq:Uint}). In Fig.~\ref{fig:UiCsCont}(a), $U/\tilde{J}$ is presented as a function of anisotropy and temperature in a contour plot. In Fig.~\ref{fig:UiCsAll}(a), $U/\tilde{J}$ is plotted as a function of temperature for various anisotropies spanning the whole Ising-like and XY-like regimes.

We see that the aforementioned discontinuous jump in zero-temperature correlations at $\Delta=1$, translates to a cusp in zero-temperature internal energy, see Fig.~\ref{fig:UiCsCont} (also see Fig.~\ref{fig:ZeroT}). We also observe a valley cusp at $\Delta=0$ for all temperatures, which is due to cusps in correlations (see Fig.~\ref{fig:CorrCont}). The internal energy increases as the anisotropy is increased from $\Delta\to-\infty$ to $\Delta=-1$; decreases in between $\Delta=-1$ and $\Delta=0$; ramps up as anisotropy is increased from $\Delta=0$ to $\Delta=1$; and declines back as the anisotropy is further increased towards $\Delta\to\infty$. For any value of the anisotropy parameter $\Delta$, the internal energy vanishes as the temperature increases (the interaction parameter $J$ decreases) as expected. In Section~\ref{sec:ZeroT} below, we further discuss the ground-state energy density $U_0/\tilde{J}$ at $T=0$. The cusps and trends can also be observed in Fig.~\ref{fig:ZeroT}, where we plot $U_0/\tilde{J}$ \emph{vs.} $\Delta$.

		\subsubsection{Specific heat}

The dimensionless specific heat $C/k$ is calculated from the internal energy density using equation~(\ref{eq:Csph}). In Fig.~\ref{fig:UiCsCont}(b), $C/k$ is presented as a function of anisotropy and temperature in a contour plot, and in Fig.~\ref{fig:UiCsAll}(b), $C/k$ is plotted as a function of temperature, for a number anisotropy parameters in between the AFM and FM Ising limits.

\begin{figure*}[t!]
	\includegraphics[scale=1.0]{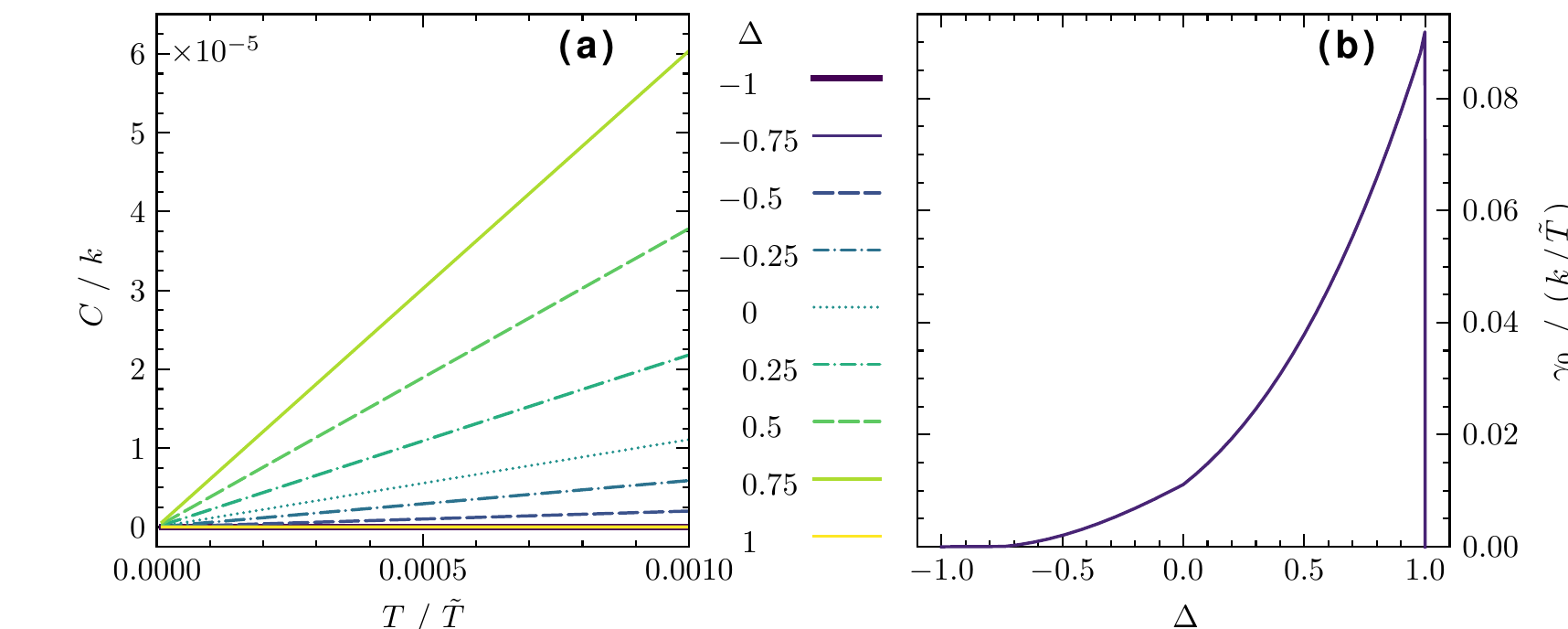}
	\caption{\label{fig:CsphLowT} Panel (a): Low-temperature specific heat $C/k$ as a function of temperature, for XY-like anisotropy parameters shown in the legend. Note that the vertical axis is scaled by a factor $10^{-5}$. We see linear behavior $C=\gamma_0\,T$ at low temperatures, see Eq.~(\ref{eq:Sommerfeld}). Panel (b): Dimensionless Sommerfeld coefficient, $\gamma_0\,\tilde{T}/k=\lim_{T\to0}[(C/k)/(T/\tilde{T})]$, as a function of anisotropy $\Delta$ in the XY-like regime, $-1\leqslant\Delta\leqslant1$. Note the cusp at $\Delta=0$ and the discontinuity at $\Delta=1$. Sommerfeld coefficient is zero at $|\Delta|=1$ and remains zero in Ising-like regimes ($|\Delta|>1$, see Fig.~\ref{fig:UiCsAll}(b)).}
\end{figure*}

The same cusps as in internal energy density also appear for the specific heat. In addition, for specific heat, cusps at critical temperatures, corresponding to FM-PM phase transitions, appear in the FM Ising-like regime ($\Delta>1$). As was shown in Section~\ref{sec:PhaseDiag}, there are no finite temperature phase transitions for the isotropic XXX models ($\Delta=\pm1$). In other regions ($\Delta<1$, $\Delta\neq1$), phase transitions are signaled not by cusps, but by maxima in specific heat as expected. In particular, specific heat peaks occurring at temperatures slightly greater than the transition temperature is a characteristic of KT transition. \cite{Loh85a} We also note the double peak structure of specific heat \emph{vs.} temperature curves for anisotropy parameter in the vicinity of $\Delta=1$. This double peak form at low temperatures was discussed in detail before, for one-dimensional XXZ model. \cite{Sariyer08} In the next section, we will discuss the low-temperature behavior of specific heat.

	\subsection{Thermodynamics at low temperatures}
		\subsubsection{Low temperature excitations} \label{sec:LowT}

Thermodynamics at low temperatures show distinct characters in XY-like ($|\Delta|<1$) and Ising-like ($|\Delta|>1$) regimes. In the XY-like regime at low temperatures, system is in algebraically ordered KT phase, where magnetization vanishes in all directions: $\langle s_i^u\rangle=0$. However, vortex-antivortex pairs are bound together in KT phase, giving rise to non-zero correlations $\langle s_i^xs_j^x\rangle>0$. The low-lying excitations are gapless, and can be understood by linear spin-wave theory or by vortex theory. \cite{Kar17} In this quantum spin-liquid phase at low-temperatures, due to a large density of low-energy states, specific heat is expected to be linear in temperature \cite{Balents10, Hirata17}:
\begin{equation}
\label{eq:Sommerfeld}
C=\gamma_0\,T ~~~ \text{for} ~~~ \left|\Delta\right|<1.
\end{equation}
Here, $\gamma_0$ is the Sommerfeld coefficient. Such linear behavior of specific heat at low-temperatures has been observed for the two-dimensional quantum spin-liquid ZnCu$_3$(OH)$_6$Cl$_2$ \cite{Helton07, Han12} and for the quasi-two-dimensional easy-plane-type XXZ ferromagnet K$_2$CuF$_4$ \cite{Hirata17}. Our results are in agreement with the expected linear form, as shown in Fig.~\ref{fig:CsphLowT}(a), where we plot specific heat as a function of temperature in XY-like regime at low-$T$.

From Fig.~\ref{fig:CsphLowT}(a), we see that the dimensionless Sommerfeld coefficient $\gamma_0\tilde{T}/k=\lim_{T\to0}[(C/k)/(T/\tilde{T})]$ is zero for $\Delta=-1$ (AFM XXX model); raises as anisotropy parameter $\Delta$ is increased; and drops back to zero at $\Delta=1$ (FM XXX model). In Fig.~\ref{fig:CsphLowT}(b), we plot the dimensionless Sommerfeld coefficient $\gamma_0\tilde{T}/k$ as a function of $\Delta$. From our numerical data, we observed that the temperature range of linearity vanishes as $\Delta\to1^-$, and in Fig.~\ref{fig:CsphLowT}(b), we see that the Sommerfeld coefficient drops to zero discontinuously at $\Delta=1$.

The vanishing Sommerfeld coefficient at $\left|\Delta\right|=1$ (see Fig.~\ref{fig:CsphLowT}(b)) is indeed an expected result. In the Ising-like regimes of $\left|\Delta\right|>1$, the low-lying excitation spectrum exhibits a gap \cite{Viswanath94}, which results in an exponential form for the specific heat (cf. Fig.~\ref{fig:UiCsAll}(b)). Hence, the Sommerfeld coefficient remains zero in Ising-like regimes as expected. Therefore, we conclude that the low-lying excitations are gapless [linear $C(T)$] in the XY-like regime of $|\Delta|<1$, while an excitation spectrum gap opens up [exponential $C(T)$] in the Ising-like regime of $|\Delta|>1$.

\begin{figure}[t!]
	\includegraphics[scale=1.0]{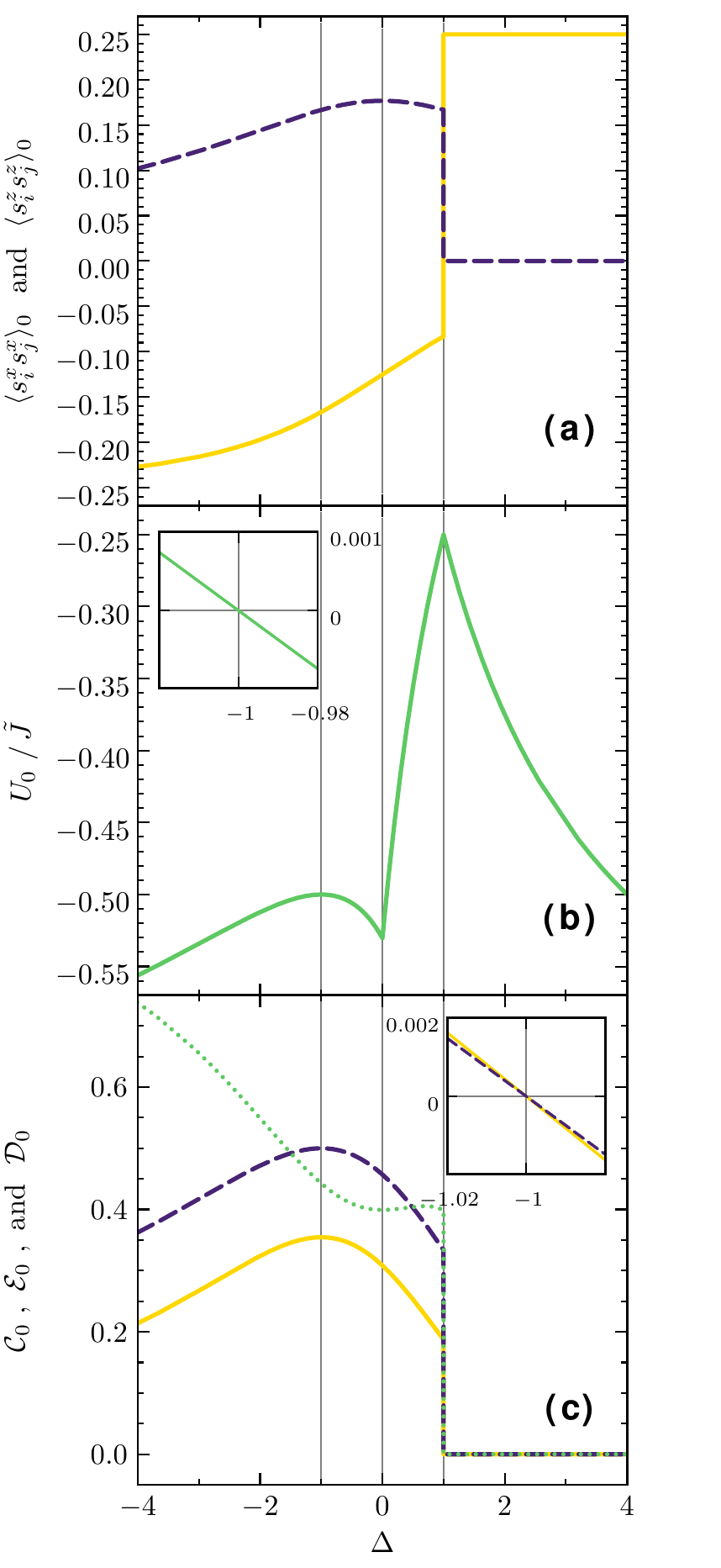}
	\caption{\label{fig:ZeroT} Physical quantities at $T=0$. Spin-spin correlations (a) $\langle s_i^zs_j^z\rangle_0$ (solid yellow) and $\langle s_i^xs_j^x\rangle_0$ (dashed purple), internal energy $U/\tilde{J}$ (b) (solid green), and entanglement measures (c) $\mathcal{C}_0$ (dashed purple), $\mathcal{E}_0$ (solid yellow), and $\mathcal{D}_0$ (dotted green) as functions of $\Delta$. Insets show the gradients with same line codes as parent functions: $d(U_0/\tilde{J})/d\Delta$ (b), $d\mathcal{C}_0/d\Delta$ and $d\mathcal{E}_0/d\Delta$ (c) as functions of $\Delta$ in the vicinity of $\Delta=-1$, where the quantities have local maxima. Thin vertical lines mark the XXX and XY models, see Table~\ref{tab:ZeroT}.}
\end{figure}

		\subsubsection{Zero-point quantum phase transitions} \label{sec:ZeroT}

In this final section of results, we discuss the zero-point ($T=0$) quantum phase transitions. We denote zero-point functions with subscripts-$0$. In Fig.~\ref{fig:ZeroT}, we plot spin-spin correlations $\langle s_i^xs_j^x\rangle_0$ and $\langle s_i^zs_j^z\rangle_0$, internal energy $U_0/\tilde{J}$ per bond, and entanglement measures $\mathcal{C}_0$, $\mathcal{E}_0$ and $\mathcal{D}_0$ as functions of anisotropy $\Delta$ at $T=0$. In fact, the RG procedure we employ, does not allow calculations at $T=0$, where at least one $J_u\to\infty$. Nevertheless, we checked that numerical results do not change for $T<10^{-6}$, and hence, we take $T=10^{-6}$ results to represent zero-point data in Fig.~\ref{fig:ZeroT}. We list the calculated values of $\langle s_i^us_j^u\rangle_0$ and $U_0/\tilde{J}$ for Ising, XY, and XXX models in Table~\ref{tab:ZeroT}.

\begin{table*}[t!]
	\caption{\label{tab:ZeroT} Zero-point spin-spin correlations and ground-state energy for Ising, XXX, and XY models. \footnote{We checked up to fourteen decimal digits that our numerical results exactly coincide with these ratios.}}
	\begin{ruledtabular}
		\begin{tabular}{c c c c c c c}
			& AFM & AFM &  & Slightly easy-plane & FM & FM \\
			& Ising & XXX & XY & FM XXX & XXX & Ising \\
			Quantity & ($\Delta\to-\infty$) & ($\Delta=-1$) & ($\Delta=0$) & ($\Delta\to1^-$) & ($\Delta=1$) & ($\Delta\to\infty$) \footnote{Correlations $\langle s_i^xs_j^x\rangle_0$ and $\langle s_i^zs_j^z\rangle_0$ take these values not only at $\Delta\to\infty$, but for the whole FM Ising-like regime ($\Delta>1$), see Fig.~\ref{fig:ZeroT}.} \\
			\hline
			$\langle s_i^xs_j^x\rangle_0$ & $0$ & $1/6$ & $\sqrt{2}/8$ & $1/6$ & $1/12$ & $0$ \\
			$\langle s_i^zs_j^z\rangle_0$ & $-1/4$ & $-1/6$ & $-1/8$ & $-1/12$ & $1/12$ & $1/4$ \\
			$U_0/\tilde{J}$ & $-3/4$ & $-1/2$ & $-3\sqrt{2}/8$ & $-1/4$ & $-1/4$ & $-3/4$
		\end{tabular}
	\end{ruledtabular}
\end{table*}

We see discontinuities in both correlations at $\Delta=1$: $\langle s_i^zs_j^z\rangle_0$ jumps from $-1/12$ at $\Delta\to1^-$, to $1/4$ (full FM correlation) at $\Delta\to1^+$; while $\langle s_i^xs_j^x\rangle_0$ jumps from $1/6$ at $\Delta\to1^-$, to $0$ (no correlations) at $\Delta\to1^+$. Both correlations take the same value $\langle s_i^zs_j^z\rangle_0=\langle s_i^xs_j^x\rangle_0=1/12$ at the isotropic XXX point $\Delta=1$ as expected. We note that both the discontinuous jumps at $\Delta=1$, and the constant correlations $\langle s_i^zs_j^z\rangle_0=1/4$ and $\langle s_i^xs_j^x\rangle_0=0$ in the whole FM Ising-like regime ($\Delta>1$) at $T=0$, has been observed before in $d=1$ \cite{Sariyer08, Justino12, Sarandy09}.

Due to the constant correlations $\langle s_i^zs_j^z\rangle_0=1/4$ and $\langle s_i^xs_j^x\rangle_0=0$, the entanglement measures $\mathcal{C}_0$, $\mathcal{E}_0$ and $\mathcal{D}_0$ all vanish in the whole FM Ising-like regime ($\Delta>1$) at $T=0$, while on the other hand, they remain finite in the AFM Ising-like regime for finite $\Delta<-1$ at $T=0$. This fact suggests that the zero-point quantum fluctuations are stronger in the AFM case, compared to the FM case.

The discontinuities in correlations actually mark the expected first-order quantum phase transition at $\Delta=1$ between KT and FM phases \cite{Orrs16, Justino12, Bishop96}. These discontinuities give rise to a cusped maximum in internal energy, and discontinuities in entanglement measures at $\Delta=1$ (see Fig.~\ref{fig:ZeroT}(b) and (c)). In addition, the Kosterlitz-Thouless quantum phase transition at $\Delta=-1$ between AFM and KT phases \cite{Orrs16, Justino12, Bishop96, You11} is signaled by maxima in internal energy, concurrence and entanglement of formation. This quantum phase transition at $\Delta=-1$ is infinite-order in $d=1$ and second-order in $d=2$. \cite{You11} The ground-state energy peak at $\Delta=-1$ was also observed by Monte Carlo simulations on square lattice \cite{Lin01}, while the concurrence peak at $\Delta=-1$ was observed in $d=2$ \cite{Gu05} as well as in $d=1$ \cite{Justino12, Sarandy09, Gu05, Dillenschneider08}. We should emphasize that the relevance of entanglement with the quantum phase transitions has been suggested by many authors before. \cite{Amico08}

We exactly obtain the expected zero-temperature (ground-state) energy per bond $U_0/\tilde{J}=-3/4$ for the AFM and FM Ising models ($|\Delta|\to\infty$), $U_0/\tilde{J}=-1/2$ and $U_0/\tilde{J}=-1/4$ for the AFM ($\Delta=-1$) and FM ($\Delta=1$) XXX models respectively (see Table~\ref{tab:ZeroT}). The ground-state energy $U_0/\tilde{J}=-3\sqrt{2}/8=-0.530331$ we calculate at $T=0$ for the XY model ($\Delta=0$) is to be compared with the square-lattice estimates $-0.403$ by variational theory \cite{Suzuki78}, $-0.405$ \cite{Hajj04} and $-0.408$ \cite{Penson80} by a real-space RG theory, $-0.411$ by coupled-cluster method \cite{Bishop96}, $-0.412$ by dressed-cluster method \cite{Hajj04}, $-0.405$ \cite{Oitmaa78, Betts77} and $-0.412$ \cite{Betts99} by finite-size extrapolations, $-0.412$ by perturbation theory \cite{Pearson77}, and $-0.407$ \cite{Loh85a, Loh85b}, $-0.412$ \cite{Lin01} and $-0.416$ \cite{Ding92a} by Monte Carlo simulations. Although there is about $20\%$ discrepancy between these square-lattice results for ground-state energy density and our result, we note that the ground-state energy per bond depends strongly on the type of lattice. \cite{Penson80, Suzuki78, Oitmaa78} The Migdal-Kadanoff RG procedure we employ works exact for classical models on a hierarchical lattice (see Fig.~\ref{fig:Hierarchical}), but not for quantum models on a square lattice. Nonetheless, it is still a good approximation for hypercubic lattices, especially for classical models. \cite{Berker79, Kaufman81, Griffiths82, Kaufman84, Erbas05}

We remind that our RG method works at its worst at the point $T=0$, $\Delta=0$. Therefore, physical quantities we calculate at other points must have less quantitative errors. We emphasize that even in the zero-temperature case, we still obtain the correct qualitative behavior, in particular, we identify the zero-point quantum phase transitions of the XXZ model.

\section{Conclusion} \label{sec:conclusion}
In conclusion, using an approximate renormalization group method, we have derived the phase diagram, critical behavior, thermodynamics, and entanglement properties of the two-dimensional uniaxially anisotropic Heisenberg model, globally at all temperatures and anisotropies. These global results are important in modeling many diverse systems such as superfluid films or magnetic monolayers in high-$T_\text{c}$ superconductors, and in understanding entanglement effects for quantum computational applications.

Nearest-neighbor spin-spin correlations, internal energy, specific heat, entanglement of formation, and quantum discord are calculated and discussed in detail. We showed that long-range-order sets in at low temperatures for all anisotropies, except for the isotropic models ($\Delta=\pm1$). We identified ferromagnetic (FM) and antiferromagnetic (AFM) phase in the Ising-like regimes, respectively for $\Delta>1$ and $\Delta<-1$; and algebraically ordered Kosterlitz-Thouless phase (KT) in the XY-like regime for $-1<\Delta<1$.

In order to calculate the magnetization components $M_u=\langle s_i^u\rangle$, we need to consider external magnetic field interactions $H_u\sum_is_i^u$ in the Hamiltonian. Calculation of (staggered-)magnetization would explicitly provide the order parameter for (A)FM phase. The order parameter for the KT phase is the helicity modulus, calculation of which would require addition of mixed interactions like $s_i^xs_j^y$ over next-nearest neighbors into the Hamiltonian. These extensions of the present work will be addressed in a future publication.

We also captured the low-temperature behavior of specific heat in the gapped (gapless) AFM and FM (KT) phases, and the quantum phase transitions at zero-temperature. The first-order quantum phase transition between FM and KT phases at $\Delta=1$ is signaled by discontinuities in spin-spin quantum correlations and hence in entanglement measures, while the second-order quantum phase transition between AFM and KT phases at $\Delta=-1$ is signaled by maxima in internal energy, concurrence and entanglement of formation.


\begin{acknowledgments}
I would like to thank Dr. A. Nihat Berker of Kadir Has University, for suggesting this problem to me and for valuable discussions; and to Dr. Aykut Erba\c{s} of Bilkent University for careful reading of the manuscript. Numerical calculations were run by a machine partially supported by the 2232 TÜBİTAK Reintegration Grant of project \# 115C135.
\end{acknowledgments}

\bibliography{XXZ2d}

\end{document}